\documentclass{aa}

\usepackage{natbib}
\bibpunct{(}{)}{;}{a}{}{,} 

\usepackage{placeins}
\usepackage{graphicx}
\usepackage{txfonts}
\newcommand{\CO}[0]{\ensuremath{^{12}\mathrm{CO}}\ }

\newcommand{\gdratio}[0]{\ensuremath{\Delta_{\rm gd}}}

\newcommand{\mdust}[0]{\ensuremath{\mathrm{M}_{\rm dust}} }
\newcommand{\rcomodel}[0]{\ensuremath{\mathrm{R_{\rm CO,\ mdl}}}}
\newcommand{\rcoobs}[0]{\ensuremath{\mathrm{R_{\rm CO,\ obs}}}}
\newcommand{\rmmmodel}[0]{\ensuremath{\mathrm{R_{\rm mm,\ mdl}}}}
\newcommand{\rmmobs}[0]{\ensuremath{\mathrm{R_{\rm mm,\ obs}}}}
\newcommand{\ratgasdustmodel}[0]{\ensuremath{\mathrm{R_{\rm CO,\ mdl}}/\mathrm{R_{\rm mm,\ mdl}}}}
\newcommand{\ratgasdustobs}[0]{\ensuremath{\mathrm{R_{\rm CO,\ obs}}/\mathrm{R_{\rm mm,\ obs}}}}
\newcommand{\ratgasdust}[0]{\ensuremath{\mathrm{R_{\rm CO}}/\mathrm{R_{\rm mm}}}}
\newcommand{\rgas}[0]{\ensuremath{\mathrm{R_{\rm CO}}}}
\newcommand{\rdust}[0]{\ensuremath{\mathrm{R_{\rm mm}}}}

\usepackage{subcaption}

\begin{document} 

    \title{Constraining the radial drift of millimeter-sized grains in the protoplanetary disks in Lupus}

   \author{L. Trapman \inst{1}
          \and
          M. Ansdell \inst{2,3}
          \and
          M.R. Hogerheijde \inst{1,4}
          \and
          S. Facchini \inst{5,6}
          \and
          C.F. Manara \inst{6}
          \and
          A. Miotello \inst{6}
          \and 
          J.P. Williams \inst{7}
          \and
          S. Bruderer \inst{5}
          }

   \institute{
            Leiden Observatory, Leiden University, Niels Bohrweg 2, NL-2333 CA Leiden, The Netherlands \\
            \email{trapman@strw.leidenuniv.nl}
        \and 
            CIPS, University of California, Berkely, 501 Campbell Hall, CA, USA
        \and
            Center for Computational Mathematics \& Center for Computational Astrophysics, Flatiron Institute, 162 Fifth Ave., New York, NY, USA
        \and
            Anton Pannekoek Institute for Astronomy, University of Amsterdam, Science Park 904, 1090 GE Amsterdam, The Netherlands
        \and
            Max-Planck-institute f\"{u}r Extraterrestrische Physik, Giessenbachstra{\ss}e, D-85748 Garching, Germany
        \and 
            European Southern Observatory, Karl-Schwarzschild-Str. 2, D-85748 Garching bei M\"unchen, German
        \and 
            Institute for Astronomy, University of Hawai‘i at M$\bar{\rm a}$noa, 2680 Woodlawn Dr., Honolulu, HI, USA
            }
   \date{Received xx; accepted yy}

  \abstract
   {Recent ALMA surveys of protoplanetary disks have shown that for most disks the extent of the gas emission is greater than the extent of the thermal emission of millimeter-sized dust. 
   Both line optical depth and the combined effect of radially dependent grain growth and radial drift may contribute to this observed effect. 
   To determine whether or not radial drift is common across the disk population, quantitative estimates of the effect of line optical depth are required. }
   {For a sample of ten disks from the Lupus survey we investigate how well dust-based models without radial dust evolution reproduce the observed \CO outer radius, and determine whether radial dust evolution is required to match the observed gas--dust size difference. }
   {Based on surface density profiles derived from continuum observations we used the thermochemical code DALI to obtain $^{12}$CO synthetic emission maps. Gas and dust outer radii of the models were calculated using the same methods as applied to the observations.
   The gas and dust outer radii $(\rgas,\rdust)$ calculated using only line optical depth were compared to observations on a source-by-source basis.}
   {For five disks, we find $\ratgasdustobs\,>\,\ratgasdustmodel$. For these disks we need both dust evolution and optical depth effects to explain the observed gas--dust size difference. For the other five disks, the observed \ratgasdust\ lies within the uncertainties on \ratgasdustmodel\ due to noise. For these disks the observed gas--dust size difference can be explained using only line optical depth effects. We also identify six disks not included in our initial sample but part of a survey of the same star-forming region that show significant (S/N $\geq3$) $^{12}$CO $J = 2\,-\,1$ emission beyond $4\times\rdust$. These disks, for which
no \rgas\ is available, likely have $\ratgasdust \gg 4$ and are difficult to explain without substantial dust evolution.
   }
   {
    Most of the disks in our sample of predominantly bright disks are consistent with radial drift and grain growth. We also find six faint disks where the observed gas--dust size difference hints at considerable radial drift and grain growth, suggesting that these are common features among both bright and faint disks. The effects of radial drift and grain growth can be observed in disks where the dust and gas radii are significantly different, while more detailed models and deeper observations are needed to see this effect in disks with smaller differences. }
   

   \maketitle

%
\section{Introduction}
\label{sec: introduction}

Over recent years the number of detected exoplanet systems has exploded, with several thousand exoplanets found around a wide range of stars. The link between these exoplanet systems and the protoplanetary disks from which they formed is still not fully understood (see e.g., \citealt{Benz2014,Morton2016}).

The behavior of the dust in protoplanetary disks is an important piece in this puzzle. In order for planets to form, dust grains have to grow from the micron sized particles in the interstellar medium to millimeter sized grains, centimeter sized pebbles, meter sized boulders, and kilometer sized planetary embryos. The rate of growth of the dust depends on both the gas and the dust surface densities, leading to radial variations (see, e.g \citealt{Birnstiel2010,Birnstiel2012}). As the grains grow, they start to decouple from the gas. As a result of gas drag, these larger dust grains lose angular momentum and start to drift inward. 
Radially dependent grain growth and inward radial drift, the combination of which we refer to here as ``dust evolution'' , together result in a decrease of the maximum grain size with distance from the star (see, e.g., \citealt{Guilloteau2011,Miotello2012,Perez2012,Perez2015,Menu2014,Tazzari2016}). 
As a consequence of dust evolution, we expect the millimeter grains to be confined in a more compact disk than the smaller grains and the gas.

The difference between the extent of the gas emission and the extent of the millimeter continuum emission has been put forward as one of the observational signatures of dust evolution. Observations almost universally show that the gas disk, traced most often by $^{12}$CO emission, is larger than the extent of the millimeter grains traced by (sub)millimeter continuum emission (see, e.g., \citealt{Dutrey1998, GuilloteauDutrey1998,Panic2008, Hughes2008,Andrews2012}). 
These observations support the idea that radial drift and grain growth are common in protoplanetary disks.

However, the observed gas--dust size difference can also be explained by the difference in optical depth between the two tracers (see also \citealt{Hughes2008}). Dust disk size (\rdust) is measured from (sub)millimeter continuum emission, which is mostly optically thin at large radii. In contrast, the size of the gas disk (\rgas) is measured using optically thick $^{12}$CO line emission. If a disk has the same radial distribution of both gas and millimeter dust, the difference in optical depth will result in an observed gas--dust size difference ($\rgas > \rdust$). Even when dust evolution has resulted in a  disk of compact millimeter grains, optical depth effects on the observed gas--dust size difference will further increase \ratgasdust. 

To quantify the relative importance of optical depth effects and dust evolution, \cite{Trapman2019} measured \ratgasdust\ for a series of models with and without dust evolution (see also \citealt{Facchini2017}). Trapman et al. found that optical depth effects alone can create gas--dust size differences of up to $\ratgasdust \simeq 4$. A gas--dust size difference $\ratgasdust \geq 4$ is a clear sign for radial drift. 
Sources showing a gas--dust size difference of $\ratgasdust \geq 4$  are rare in current observations, but one example is CX Tau, for which \cite{Facchini2019} measured $\ratgasdust \simeq 5$.

With the advent of the Atacama Large Millimeter/sub-Millimeter Array (ALMA) it has become possible to do surveys of entire star-forming regions, taking $\sim1$ minute snapshots of each disk at moderately high resolution ($\sim0{\farcs}25-0{\farcs}40$). This has allowed us to study the properties of the complete disk population (e.g., Taurus, \citealt{Andrews2013,WardDuong2018,Long2018,Long2019}; Lupus, \citealt{ansdell2016,ansdell2018}; Chamaeleon I, \citealt{Pascucci2016,Long2017}; $\sigma$-Ori \citealt{ansdell2017}; Upper Sco, \citealt{Barenfeld2016,barenfeld2017}; Corona Australis, \citealt{Cazzoletti2019};  and Ophiuchus, \citealt{Cox2017,Cieza2019,Williams2019}). 
In the survey of the Lupus star forming region, \cite{ansdell2018} measured the gas and dust outer radii (R$_{\rm gas}$, R$_{\rm dust}$) for a sample of 22 disks. These latter authors found that the extent of the gas exceeds the extent of the dust, with an average ratio of $\mathrm{R_{\rm gas}}/\mathrm{R_{\rm dust}} =  1.96 \pm 0.04|_{\sigma_{\rm obs}} \pm 0.57|_{\sigma_{\rm dispersion}}$. Here, $\sigma_{\rm obs}$ is the uncertainty due to the errors on the observed outer radii, whereas $\sigma_{\rm dispersion}$ is the standard deviation of the sample.

The average gas--dust size difference $\langle \ratgasdust \rangle = 1.96$ is much lower than the value of $\ratgasdust \sim 4$ found by \cite{Trapman2019} to be a clear indication of dust evolution. This would suggest that almost none of the 22 disks show signs of having undergone dust evolution. However, low $^{13}$CO and C$^{18}$O line fluxes observed for these sources indicate that they also have a low CO content, which lowers the contribution of optical depth effects to the gas--dust size difference. A more detailed analysis is required to determine whether the \ratgasdust\ observed for the disks in Lupus is a sign of radial drift and grain growth or can be reproduced using only optical depth effects. 

In this work, the gas structure of a sample of ten disks taken from the Lupus survey is modeled using the thermochemical code DALI \citep{Bruderer2012,Bruderer2013} under the assumption that gas and dust follow the same density structure.
The resulting \ratgasdust, set only by optical depth, are compared to observations on a source-by-source basis, and conclusions are drawn on whether or not dust evolution, that is, radial drift and radially dependent grain growth, is needed to match the observed \ratgasdust.

Section \ref{sec: observations and sample selection} describes the observations and sample selection. The models are described in Section \ref{sec: methods} and we describe how the gas and dust outer radii are measured. In Section \ref{sec: results} the gas models are compared to the observations in terms of the extent of the gas as traced by $^{12}$CO and the gas--dust size difference. The role of noise in measuring \ratgasdust\ is also discussed. In Section \ref{sec: discussion} we examine the Lupus disks with unresolved dust emission that were detected in $^{12}$CO and discuss whether \ratgasdust\ could be larger for more compact dust disks. 


\section{Observations and sample selection}
\label{sec: observations and sample selection}

\subsection{Observations}
\label{sec: observations}

The disks analyzed in this paper are a subsample of the ALMA Lupus disk survey \citep{ansdell2016,ansdell2018} (id: ADS/JAO.ALMA\#2013.1.00220.S, Band 7, and ADS/JAO.ALMA\#2015.1.00222.S, Band 6) and the Lupus Completion Survey (id: ADS/JAO.ALMA\#2016.1.01239.S Band 6 and 7). 

The band 7 observations were taken with an array configuration covering baselines between 21.4 and 785.5 m. The resulting average beamsize for the continuum is $0\farcs34 \times 0\farcs28$. The bandwidth-weighted mean continuum frequency was 335.8 GHz (890 $\mu$m). The spectral setup included two windows covering the $^{13}$CO $J = 3\,-\,2$ and C$^{18}$O $J = 3\,-\,2$ transitions centered at 330.6 GHz and 329.3 GHz, respectively. Both windows have channel widths of 0.122 MHz, corresponding to a velocity resolution of 0.11 km s$^{-1}$. Further details on the observational setup and data reduction can be found in \cite{ansdell2016}.

The targets of the observations consist of a sample of sources selected from the Lupus star-forming complex (clouds I to IV) that were classified as Class II or Flat IR spectra disks \citep{Merin2008}.  The sample totaled 93 objects of which 61 were detected in the continuum at $\geq 3 \sigma$. The ALMA observations are complemented by a VLT/X-shooter spectroscopic survey by \cite{alcala2014,alcala2017}. These latter authors derive fundamental stellar parameters for the Class II objects of the region.

The Band 6 observations were taken with a more extended configuration, covering baselines between 15 and 2483 m. As a result, the average beam size for the continuum is $0\farcs25 \times 0\farcs24$, slightly smaller than the one for the Band 7 observations. The bandwidth-weighted mean continuum frequency of these observations was 225.66 GHz (1.33 mm). Three windows were included in the spectral setup, covering the $^{12}$CO $J = 2\,-\,1$, $^{13}$CO $J = 2\,-\,1$ and C$^{18}$O $J = 2\,-\,1$ transitions centered at 230.51, 220.38, and 219.54 GHz respectively. Each spectral window has a bandwidth of 0.12 GHz, a channel width of 0.24 MHz, and velocity resolution of 0.3 km s$^{-1}$. More details of the observations can be found in \cite{ansdell2018}. 
We note that the sample in \cite{ansdell2018} covered four additional sources while also excluding two sources later found to be background red giants \citep{Frasca2017}. Neither of these changes affect our sample selection.

\subsection{Sample selection}
\label{sec: sample selection}

In total, 48 of the 95 targets were detected both in 1.33 mm continuum and $^{12}$CO $J = 2-1$ line emission. For 22 of these sources, the signal-to-noise ratio (S/N) in the channel maps was high enough to measure the gas outer radius defined as the radius enclosing 90\% of the $^{12}$CO flux \citep{ansdell2018}. For our sample, we select 10 of these 22 disks that have dust surface density profiles derived by \cite{Tazzari2017}, which is a prerequisite for our analysis. IM Lup formally meets our selection criteria, but it is excluded due to its structural complexity.

Of the remaining 11 disks with an observed gas outer radius that were not included in our sample, two sources were covered by the Lupus completion survey (ID: 2016.1.01239.S, PI: van Terwisga) and were not included in the analysis by \cite{Tazzari2017}. The remaining 9 were also not included in the analysis by \cite{Tazzari2017} due to the presence of a clear cavity in the image plane. We note that several other disks in the sample (e.g., Sz 84, Sz 100, MY Lup) have been identified as transition disks, either in the higher resolution band 6 observations or in the visibilities (cf. \citealt{Tazzari2017,vdMarel2018}). We also excluded Sz 73 from our analysis as it was not detected in $^{13}$CO, preventing us from calibrating its CO content.

Our final sample consists of ten disks (in order of decreasing dust mass): Sz 133, Sz 98,  MY Lup, Sz 71, J16000236-4222115, Sz 129, Sz 68, Sz 100, Sz 65 and Sz 84. Their properties are shown in Table \ref{tab: source properties}.

\begin{table*}[htb]
\centering
\caption{\label{tab: source properties} Source properties}
\begin{tabular*}{0.9\textwidth}{l|llll|lclllrr}
\hline\hline
 & \multicolumn{4}{c}{Stellar properties$^{1,\dagger}$}  & \multicolumn{7}{c}{Disk properties$^{2,3,\dagger}$} \\
Name & L$_*$ & $T_{\rm eff}$ & M$_*$ & $d$ & $\gamma$ & M$_{\rm dust}$ & R$_{\rm c}$ & PA & $i$ & R$_{\rm dust}$ & R$_{\rm gas}$\\
 & (L$_{\odot}$) &  (K)  & (M$_{\odot}$) &  (pc) & $  $ & ($\times10^{-4} \mathrm{M}_{\odot}$) & (AU) & (deg) & (deg) & (AU) & (AU)\\
\hline
Sz 133 & 0.07 & 4350 & 0.63 & 153 & -0.17 & 2.9 & 68.1 & 126.29 & 78.53 & 145.9 & 225.5 \\
Sz 98 & 1.53 & 4060 & 0.67 & 156 & 0.11 & 2.8 & 155.4 & 111.58 & 47.1 & 148.4 & 279.6 \\ 
MY Lup & 0.85 & 5100 & 1.09 & 156 & -0.59 & 2.8 & 63.3 & 58.94 & 72.98 & 114.8 & 204.6 \\ 
Sz 71 & 0.33 & 3632 & 0.41 & 155 & 0.25 & 2.6 & 88.0 & 37.51 & 40.82 & 98.7 & 229.7 \\ 
J16000236 & 0.18 & 3270 & 0.23 & 164 & -0.2 & 2.6 & 98.1 & 160.45 & 65.71 & 122.6 & 301.0 \\ 
Sz 129 & 0.43 & 4060 & 0.78 & 161 & -0.33 & 2.5 & 54.2 & 154.94 & 31.74 & 73.3 & 141.2 \\ 
Sz 68 & 5.42 & 4900 & 2.13 & 154 & -0.39 & 1.3 & 14.4 & 175.78 & 32.89 & 39.1 & 73.0 \\
Sz 100 & 0.08 & 3057 & 0.14 & 136 & -1.52 & 0.6 & 41.1 & 60.2 & 45.11 & 56.1 & 121.9 \\
Sz 65 & 0.89 & 4060 & 0.7 & 155 & 0.12 & 0.5 & 29.0 & 108.63 & 61.46 & 67.3 & 191.5 \\ 
Sz 84 & 0.13 & 3125 & 0.17 & 152 & -0.98 & 0.4 & 41.3 & 167.31 & 73.99 & 81.4 & 148.6 \\ 
\hline
\end{tabular*}
\captionsetup{width=.86\textwidth}
\caption*{\footnotesize{$^{1}$: \cite{alcala2014,alcala2017}. $^{2}$ Dust and gas radii from \cite{ansdell2018}. $^{3}$: Other disk parameters from \cite{Tazzari2017} 
$^{\dagger}$: Both stellar and disk parameters were recalculated using the Gaia DR2 distances ( \citealt{GaiaDR2_2018, BailerJones2018}; cf. Appendix A in \citealt{Manara2018}) and Appendix A in \citealt{Alcala2019}}}
\end{table*}

\begin{figure}
     \centering
    \includegraphics[width=0.9\columnwidth]{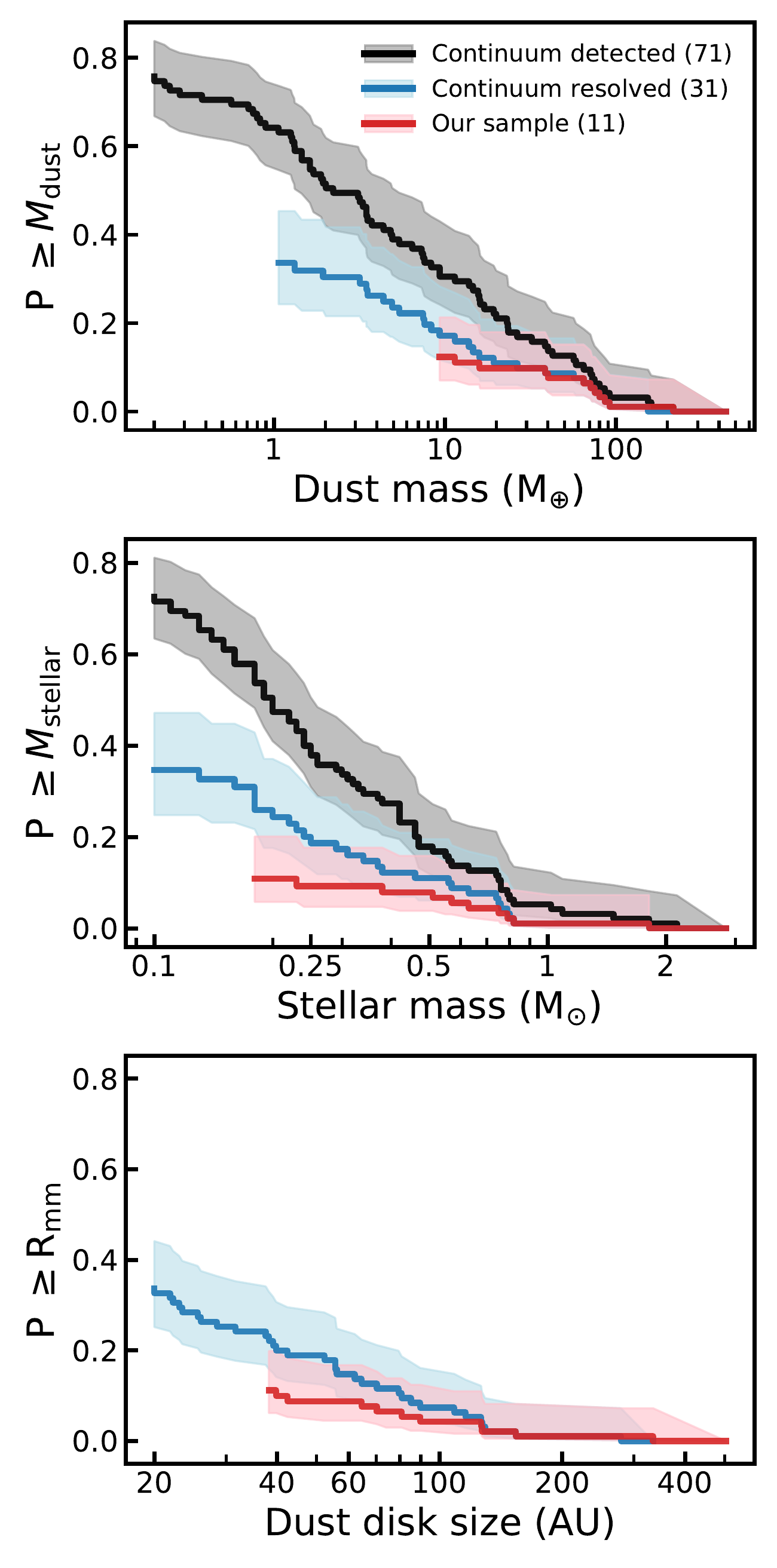}
    \caption{\label{fig: sample properties}  Cumulative distribution of our sample in relation to the full Lupus disk population. 
    \textbf{Top panel:} Dust masses taken from \citealt{ansdell2016} of all Lupus disks detected in continuum (gray), Lupus disks resolved in continuum (blue) and our sample (red; see Section \ref{sec: sample selection}). 
    \textbf{Middle panel:} Same as above, but showing stellar masses derived from X-SHOOTER spectra \cite{alcala2017}, but recalculated using the new Gaia DR2 distances (see Appendix A in \citealt{Alcala2019})
    \textbf{Bottom panel:} Same as above, but showing dust outer radii $(\rdust)$.}
\end{figure}

Figure \ref{fig: sample properties} shows the comparison between our sample and the full survey \citep{ansdell2016,ansdell2018}. We note that our sample is biased towards the most massive disks (in dust). This is likely due to the fact that both resolving the dust and being able to measure the gas outer radius biases the sample to the brightest, most easily detected and therefore most massive disks.

\section{Methods}
\label{sec: methods}

The observed difference in extent between gas and dust (gas--dust size difference), quantified by the ratio of the radii enclosing 90\% of the $^{12}$CO $J=2\,-\,1$ and 1.3 millimeter emission (\ratgasdustobs), is set by a combination of line optical depth effects and dust evolution, that is, radial drift and radially dependent grain growth. 
We can use \ratgasdustobs\ to identify whether or not a disk has undergone dust evolution provided that we know the contribution of optical depth to the gas--dust size difference. To find out if the disks in Lupus show signs of dust evolution, our approach is the following. Based on observational constraints, we set up source-specific models for the ten disks in our sample, where we assume that dust evolution has not occurred. We use the thermochemical code \texttt{DALI} \citep{Bruderer2012,Bruderer2013} to create synthetic dust continuum and CO line emission maps. Gas and dust outer radii of the model (\rcomodel,\rmmmodel) are measured from the emission using the same methods that were applied to the observations. Combining \rcomodel\ and \rmmmodel, we calculate \ratgasdustmodel, which for our models is only based on optical depth effects. In this context, sources for which \ratgasdustmodel \   is smaller than \ratgasdustobs \ would indicate that some combination of radial drift and grain growth has occurred. 

We should note here that both radial drift and grain growth produce a similar radial distribution of dust grain sizes, that is, that larger grains are concentrated closer to the star, and therefore these two effects lead to a similar observed \ratgasdustobs. \cite{BirnstielAndrews2014} showed that a sharp outer edge of the dust emission is a clear signature of radial drift. Unfortunately our observations lack the sensitivity to detect this sharp edge. Throughout this work we therefore use the term ``dust evolution'' to refer to the combined effect of radial drift and grain growth.

\subsection{DALI}
\label{sec: dali}

To calculate CO line fluxes and produce images, we use the thermochemical code {D}ust {A}nd {LI}nes (\texttt{DALI}; \citealt{Bruderer2012,Bruderer2013}). \texttt{DALI} takes a physical 2D disk model and calculates the thermal and chemical structure self-consistently. Using the stellar spectrum, the UV radiation field inside the disk is calculated. The computation is split into three steps. At the start the dust temperature structure and the internal radiation field are calculated by solving the radiative transfer equation using a 2D Monte Carlo method. For each point, the abundances of the molecular and atomic species are calculated by solving the time-dependent chemistry. The excitation levels of the atomic and molecular species are computed using a nonlocal thermodynamic equilibrium (NLTE) calculation. Based on these excitation levels, the gas temperature can be calculated by balancing the heating and cooling processes. Since both the chemistry and the excitation depend on temperature, an iterative calculation is used to find a self-consistent solution. Finally, the model is ray-traced to construct spectral image cubes and line profiles. A more detailed description of the code can be found in Appendix A of \cite{Bruderer2012}. 

\subsubsection{Chemical network}
\label{sec: chemical network}

We use the CO isotopolog chemical network from \cite{Miotello2014} which includes both CO freeze-out and photodissociation of $^{12}$CO and its $^{13}$CO, C$^{17}$O, and C$^{18}$O isotopologs individually. This is an extension of the standard chemical network in DALI, which is based on the UMIST 06 network \citep{woodall2007,Bruderer2012,Bruderer2013}. Reactions included in the network are H$_2$ formation on the grains, freeze-out, thermal and non-thermal desorption, hydrogenation of simple species on ices, gas-phase reactions, photodissociation, X-ray- and cosmic-ray-induced reactions, polycyclic aromatic hydrocarbon (PAH) grain charge exchange and/or hydrogenation, and reactions with vibrationallly excited H$_2^*$. The implementation of these reactions can be found in Appendix A.3.1 of \cite{Bruderer2012}. \cite{Miotello2014} expanded the chemical network to include CO isotope-selective processes such as photodissociation (see also \citealt{Visser2009}).

\subsection{The physical model}
\label{sec: the physical model}

For the surface density of the model, we use the surface density profiles from \cite{Tazzari2017}. These latter authors fitted the 890 $\mu$m visibilities of each source in our sample using a simple disk model with a tapered power law surface density and a two-layer temperature structure (see \citealt{ChiangGoldreich1997}). The tapered power law surface density is given by \cite{LyndenBellPringle1974, Hartmann1998}

\begin{equation}
    \label{eq: surface density}
    \Sigma = \frac{\left(2-\gamma\right) M_{\rm disk}}{2\pi R_{\rm c}^2} \left(\frac{R}{R_{\rm c}}\right)^{-\gamma} \exp\left[-\left( \frac{R}{R_{\rm c}} \right)^{2-\gamma} \right],
\end{equation}
where $M_{\rm disk}$ is the total disk mass, $\gamma$ is the slope of the surface density, and $R_{\rm c}$ is the characteristic radius of the disk.

By using this surface density for the gas in our model, the gas and millimetre-sized grains follow the same surface density profile. Our null hypothesis is thus that no radial-dependent dust evolution has occurred. 

\subsubsection{Vertical structure}
\label{sec: vertical structure}

\begin{figure}
    \centering
    \includegraphics[width=0.8\columnwidth]{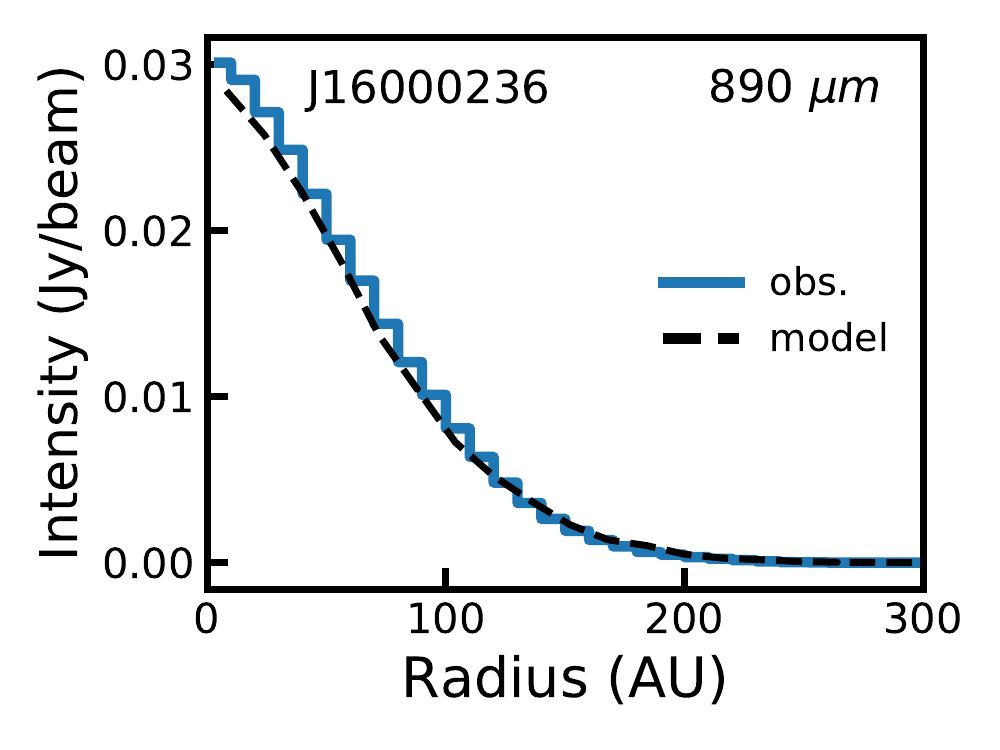}
    \caption{\label{fig: 890um continuum comparison example} Example comparison between model and observed 890 $\mu$m continuum intensity profile for J16000236-4222115. Similar comparisons for the other disks can be found in Figure \ref{fig: 890um continuum comparison}.}
\end{figure}

In their original fit, \cite{Tazzari2017} use a two-layer vertical structure to calculate the temperature structure of their models (see \citealt{ChiangGoldreich1997}). This vertical structure consists of a thin upper layer that intercepts and is superheated by the stellar radiation. This upper layer re-emits in the infrared and heats the interior of the disk. 
To correctly calculate the CO chemistry and emission, we instead need to run a full two-dimensional physical-chemical model. Assuming hydrostatic equilibrium and a vertically isothermal disk, the vertical density structure is given by a Gaussian distribution

\begin{equation}
\label{eq: vertical structure}
n(R,z) =  \frac{1}{\sqrt{2 \pi}}\frac{1}{H} \exp \left[ -\frac{1}{2}\left(\frac{z}{H}\right)^2\right],
\end{equation}
where $H = Rh$ is the physical height of the disk and the scale height $h$ is parametrized by
\begin{equation}
\label{eq: scale height}
h = h_c \left(\frac{R}{R_c}\right)^{\psi}.
\end{equation}
Here, $h_c$ is the scale height at $R_c$ and $\psi$ is known as the flaring angle. In this work, $(h_c, \psi) = (0.1,0.1)$ is assumed for all disks.

Compared to the models fitted by \cite{Tazzari2017}, our models have a different vertical structure and their temperature structure is calculated differently (cf. Section \ref{sec: dali}). It is therefore worthwhile to confirm that our models still reproduce the observed 890 $\mu$m continuum emission. As an example, Figure \ref{fig: 890um continuum comparison example} compares the model and observed 890 $\mu$m radial intensity profile for J16000236\,-\,4222115. As shown in the figure, our model matches the observation. Similar figures for all ten disks are shown in Figure \ref{fig: 890um continuum comparison}. Except for Sz\,133, our models reproduce the observed 890 $\mu$m radial intensity profile.

\subsubsection{Total CO content}
\label{sec: total co content}

The gas outer radius is measured from $^{12}$CO emission and increases with the total CO content (see, e.g., \citealt{Trapman2019}). Observations revealed low $^{13}$CO and C$^{18}$O line fluxes for most disks in Lupus, indicating that they have a low total CO content (see, e.g., \citealt{ansdell2016,ansdell2018,miotello2017}). 

Two explanations have been suggested to explain the low CO isotopolog line fluxes. CO isotopologs $^{13}$CO and C$^{18}$O are often used to measure the total gas mass. The low $^{13}$CO and C$^{18}$O line fluxes could indicate that these disks have low gas masses, suggesting low gas-to-dust mass ratios (\gdratio) of the order of $\gdratio \simeq 1-10$.
 
Alternatively, the low $^{13}$CO and C$^{18}$O line fluxes could be due to an overall underabundance of volatile CO. In this case the disks do not have a low gas mass, but instead some process not currently accounted for has removed CO from the gas phase. Several processes have been suggested to explain the underabundance of CO. One possibility is linked to grain growth, where CO freezes out and becomes locked up in larger bodies, preventing it from re-entering the gas-phase chemistry (see, e.g., \citealt{Bergin2010,Bergin2016,Du2015,Kama2016}). 
Alternatively, CO can be removed by converting it into more complex organics such as CH$_3$OH that have higher freeze-out temperatures or turning it into CO$_2$ and/or CH$_4$ ice (see, e.g., \citealt{Aikawa1997, Favre2013, Bergin2014, Bosman2018, Schwarz2018}). We note that neither of these processes is included in \texttt{DALI}. 
Recent C$_2$H observations in a subsample of Lupus disks are in agreement with this second hypothesis, that is, with C and O being underabundant in the gaseous outer disk \citep{Miotello2019}.

It should be noted here that the two explanations for the low total CO content discussed here have very different implications for the evolution of dust in the disk. If the low CO fluxes are indicative of low gas-to-dust mass ratios, 
dust grains will not be well coupled to the gas, increasing the effects of fragmentation and thus limiting the maximum grain size in the disk.
If disks are underabundant in CO but have $\gdratio\sim100,$ dust grains stay well coupled to the gas for longer and grow to larger sizes, and for most of the disks radial drift will set the maximum grain size.

For each source in our sample we examine both scenarios. We run a gas depleted disk model, where \gdratio\ is lowered until the model reproduces the observed $^{13}$CO 3\,-\,2 line flux. In addition, we run a CO underabundant model for each source, where we lower the C and O abundances until the $^{13}$CO 3\,-\,2 line flux matches the observations. Our tests show that both approaches yield near identical results. In this work we therefore only show the CO underabundant models.

\subsubsection{Dust properties}
\label{sec: dust settling}

Dust settling is included parametrically in the model by splitting the grains into two populations:
\begin{itemize}
    \item Small grains (0.005-1 $\mu$m) are included with a (mass) fractional abundance $1-f_{\rm large}$ and are assumed to be fully mixed with the gas.
    \item Large grains (1-$10^3\ \mu$m) are included with a fractional abundance $f_{\rm large}$. To simulate the large grains settling to the midplane, these grains are constrained to a vertical region with scale height $\chi h;\ \chi < 1$.
\end{itemize}
The opacities are computed using a standard interstellar medium (ISM) dust composition following \cite{WeingartnerDraine2001}, with a MNR \citep{mathis1977} grain size distribution between the grain sizes listed above.

In our models, we set $f_{\rm large} = 0.99$ and $\chi =0.2$, thus assuming that the majority of the dust mass is in the large grains that have settled to the midplane of the disk.

We note that in our analysis we keep the disk flaring structure and dust settling fixed and identical for all ten sources in our sample. In practice these parameters will likely vary between different disks. In Appendix \ref{app: flaring and dust settling} we show that varying these parameters changes \rcomodel\ by less than 10 \%.
    
\subsubsection{Stellar spectrum}
\label{sec: stellar spectrum}
\cite{alcala2014,alcala2017} used VLT X-Shooter spectra to derive stellar properties for all our sources.  Using their stellar luminosity and effective temperature, re-scaled to account for the new Gaia DR2 distances (see also Appendix A in \citealt{Alcala2019}), we calculate the blackbody spectrum to use as the stellar spectrum for each source. Excess UV radiation that is expected as a result of accretion is added to the stellar spectrum as blackbody emission with T = 10000 K. The total luminosity of this component is set to the observed accretion luminosity \citep{alcala2017}. For three sources (Sz 65, Sz 68 and MY Lup) only an upper limit of the accretion luminosity is known. We use this upper limit as the accretion luminosity in the models for these sources.

Some of the sources in our sample have anomalously low stellar luminosities, which could be linked to the high inclination of their disk. The main consequence of underestimating the stellar luminosity will be that our disk model is too cold. Increasing the stellar luminosity will have a very similar effect to increasing the flaring of the disk, as both increase the temperature in the disk. In Appendix \ref{app: flaring and dust settling} we show that increasing the flaring has only a minimal effect on the gas outer radius and therefore we do not expect an underestimation of the stellar luminosity to affect our results.

\subsection{Measuring model outer radii}
\label{sec: measuring model outer radii}
\begin{figure}
    \centering
    \includegraphics[width=\columnwidth]{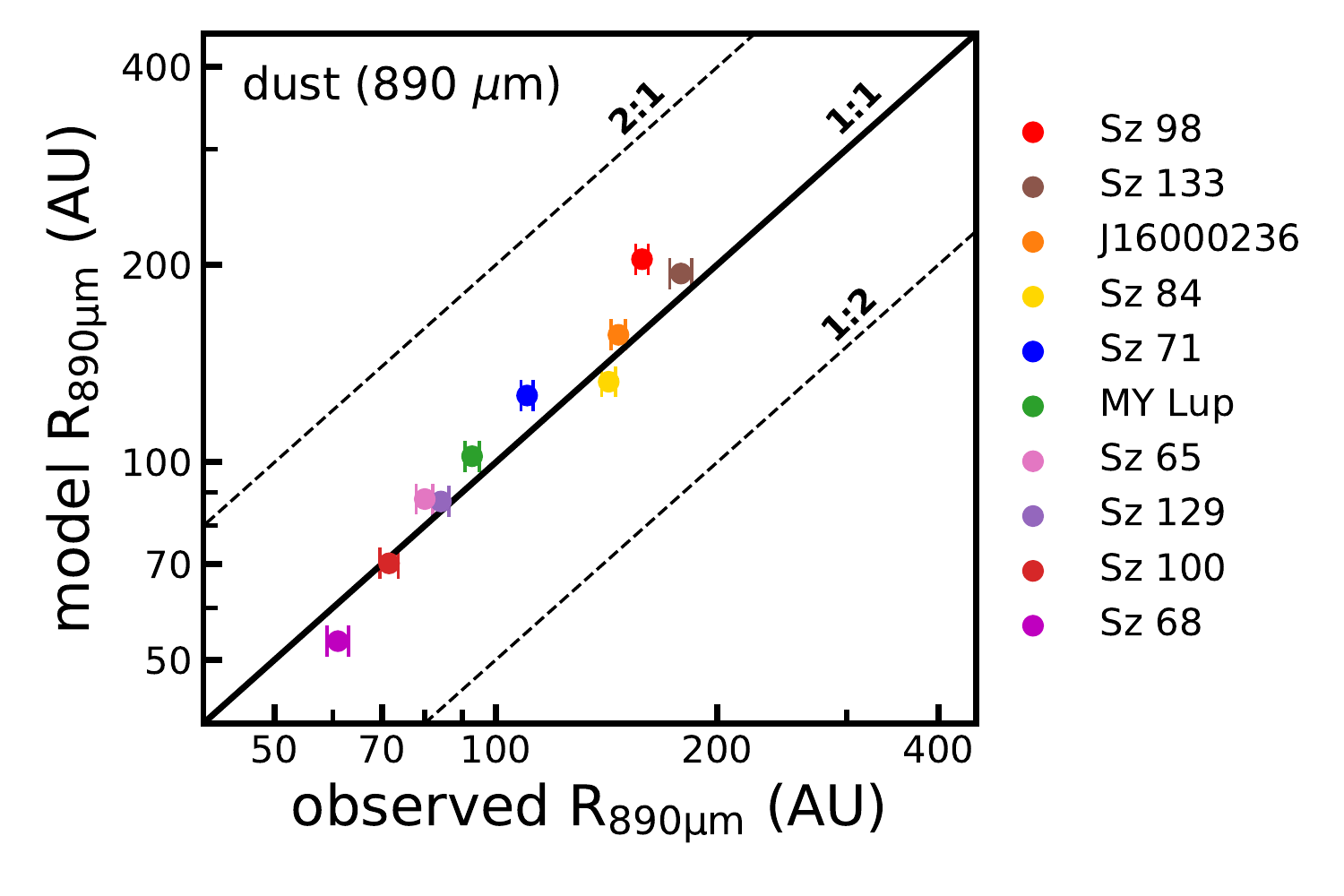}
    \caption{\label{fig: 890um dust size comparison} Comparison between model and observed dust outer radii based on the 890 $\mu$m continuum emission. Differences are within 15\% with the exception of Sz 98 (30\%).}
\end{figure}

We follow the same approach as \cite{ansdell2018} to measure the gas and dust outer radii from our models. These latter authors define the outer radius as that which encloses 90 \% of the total flux. In both the observations the gas outer radius (\rgas) is measured from the $^{12}$CO 2\,-\,1 line emission and the dust outer radius (\rdust) is measured from the 1.3 millimeter continuum emission. The outer radius is measured using a curve of growth method where the flux is measured in increasingly larger elliptical apertures until the measured flux reaches 90 \% of the total flux. The inclination $(i)$ and position angle (PA) of the apertures are chosen to match the orientation of the continuum emission (see \citealt{Tazzari2017}). 
A Keplerian masking technique is applied to the line emission to increase the S/N of the CO emission in the outer parts of the disk (for details, see Appendix \ref{app: keplerian masking}).
Uncertainties on \rgas\ and \rdust\ are determined by taking the uncertainties on the total flux and using the curve of growth method to propagate these uncertainties into the observed outer radius. 

\begin{figure*}[!th]
    \centering
    \includegraphics[width=0.9\textwidth]{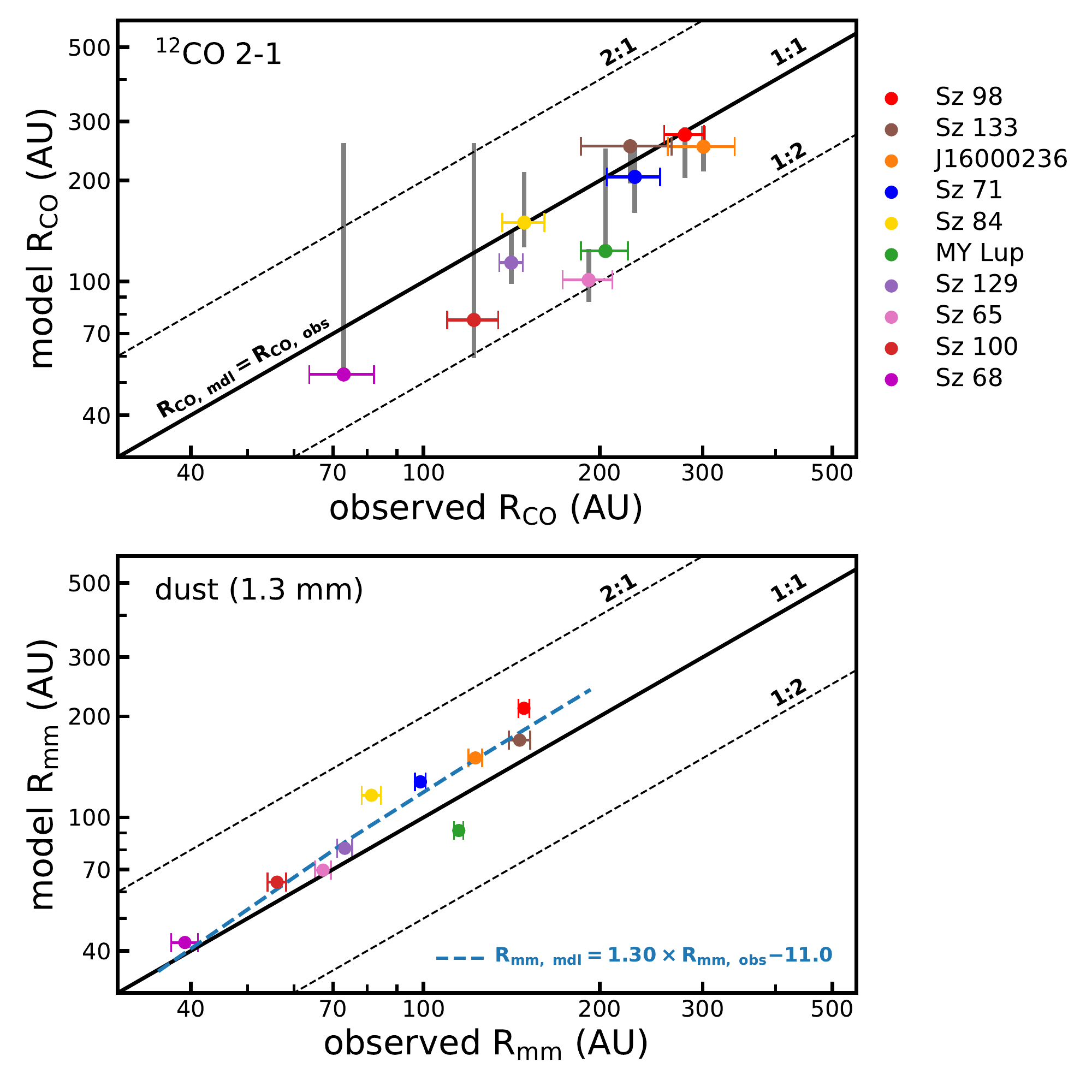}
    \caption{\label{fig: gas and dust outer radii}
    Disk size comparison of the models and the observations. 
    \textbf{Top panel:}  Gas disk size (\rgas), defined as the radius enclosing 90\% of the $^{12}$CO $J = 2\,-\,1$ flux. Observed \rgas, including uncertainties, are shown as a colored error bar \citep{ansdell2018}. Gray vertical lines denote the range of \rcomodel\ measured after noise is included (see Appendix \ref{app: adding noise to the model}). The upper and lower points of the gray line show the 84$^{\rm th}$ and 16$^{\rm th}$ quantiles, respectively, of the noisy R$_{\rm CO}$ distribution (cf. Appendix \ref{app: noisy rgas distributions}).
    \textbf{Bottom panel:} As above, but showing the dust disk size, defined as the radius that encloses 90\% of the 1.3 mm continuum flux. The blue dashed line shows the best fit of the offset between \rmmmodel\ and \rmmobs\ (see Section \ref{sec: gas-dust size ratios}).
    }
\end{figure*}

We measure \rgas\ and \rdust\ from our models following the same procedure. First, our models are raytraced at the observed inclination and the resulting synthetic $^{12}$CO line emission cubes and 1.3 millimeter continuum emission maps are convolved with the beam of the Band 6 observations (see Section \ref{sec: observations}; \citealt{ansdell2018}). 
We note that this approach is a simplification of producing full synthetic observations of our models by sparsely sampling the Fourier transform of our model and reconstructing it in the image plane with the CLEAN algorithm. However, tests show that both methods yield approximately the same measured gas and dust outer radii.

For the CO, we add noise to the synthetic $^{12}$CO $J = 2\,-\,1$ spectral cube following the procedure outlined in Appendix \ref{app: adding noise to the model} and we apply the Keplerian mask that was used for the observations before collapsing the spectral cube along the spectral axis to create a moment-zero map. The outer radii are measured using the curve of growth method, also using $i$ and PA of the continuum for the elliptical apertures. Figure \ref{fig: 890um dust size comparison} shows a comparison between model and observed outer radii based on the 890 $\mu$m continuum emission. The models and observations agree to within 15 \% except for Sz 98 ($\sim 29\%$). Figures \ref{fig: 890um continuum comparison example} and \ref{fig: 890um dust size comparison} show that our models are able to reproduce the continuum intensity profile and the extent of the 890 $\mu$m continuum observations. 

\section{Results}
\label{sec: results}

\subsection{Observed versus modeled disk sizes}
\label{sec: observed versus modelled disk sizes}

Having measured the gas and dust outer radii of our models, we can compare them to the observations. 
In Figure \ref{fig: gas and dust outer radii} we compare the gas and dust outer radii of the model (\rcomodel, \rmmmodel) to the observed gas and dust outer radii (\rcoobs,\rmmobs).

The top panel of Figure \ref{fig: gas and dust outer radii} shows that all models are either equal in size to or smaller than the observations in terms of the measured gas outer radius. If we account for the uncertainty on \rcoobs\, shown in the figures as the colored error bar, we find that for six of the ten models 
$(\sim 60\%)$  the modeled gas outer radius (\rcomodel) is smaller than the observed gas outer radius (\rcoobs). These are predominantly the smaller disks $(\rcoobs \lessapprox 200\ \mathrm{AU})$. 

The gray bars in Figure \ref{fig: gas and dust outer radii} show the range of \rcomodel\ that we measure after adding noise to the synthetic $^{12}$CO $J = 2\,-\,1$ spectral cube following the procedure outlined in detail in Appendix \ref{app: adding noise to the model}. Briefly, we take a noise map from the observed spectral cube at random and add it to the model spectral cube, apply the Keplerian mask, and measure \rcomodel\ using the curve of growth method outline in Section \ref{sec: measuring model outer radii}. This process is repeated for approximately $1000$ noise realisation and gives us a distribution of possible \rcomodel\ that could be measured from our models in the presence of noise (see Figure \ref{fig: noisy realisation example}). 

For most of our sources we see that the uncertainty on \rcomodel, which is represented by the range of noisy \rcomodel, is smaller than or similar to the uncertainty on \rcoobs. 
This shows that propagating the uncertainty in the total flux through the curve of growth into the observed outer radius is in most cases a conservative estimate of the effect of noise on measuring the outer radius.  

The range of noisy \rcomodel\ is the largest of the two smallest disks in the sample, Sz 68 and Sz 100 (see Figure \ref{fig: gas and dust outer radii}). Within our sample, these disks are also among the faintest in $^{12}$ CO $J = 2\,-\,1$ emission. The compact, faint emission makes their curve of growth more susceptible to the effects of noise. These results show that care should be taken when measuring the gas disk size of faint, compact disks using a curve of growth method. 

The bottom panel of Figure \ref{fig: gas and dust outer radii} compares modeled dust outer radius (\rmmmodel) and observed dust outer radius (\rmmobs), both measured from the 1.3 millimeter continuum emission.
In the figure we can see that the \rmmmodel\ of four disk models are significantly larger than \rmmobs: namely Sz\,84 (43\%), Sz\,71 (29\%), J16000236 (23\%), and Sz\,98 (42\%), and the disk model for MY Lup is significantly smaller (20\%).
The differences of \rdust\ between the model and the observations are larger than the differences seen at $890\,\mu$m (cf. Figure \ref{fig: 890um dust size comparison}). 
This indicates that our models overproduce the extent of the 1.3 millimeter continuum emission despite reproducing the extent of the continuum emission at $890\,\mu$m.
A potential explanation for this is the lack of dust evolution in our models. Based on dust evolution, the larger 1.3 millimeter grains should be more concentrated toward the star compared to the $890\,\mu$m grains, which should be reflected by dust outer radii decreasing with wavelength (see, e.g., \citealt{Tripathi2018}). In our models we do not take this into account. Instead, we
base our models on the $890\,\mu$m continuum emission, and thus the extent of the $890\,\mu$m grains, and assume the same radial extent for the 1.3 millimeter grains. The fact that \rmmmodel\ is larger than \rmmobs\ would therefore suggest that without including dust evolution we are overestimating the radial extent of the 1.3 millimeter grains. 

\subsection{Gas--dust size difference: models versus observations}
\label{sec: gas-dust size ratios}

The ratio between \rgas\ and \rdust\ of a disk encodes whether it has undergone dust evolution. From the gas and dust outer radii of the models we can calculate their gas--dust size ratio (\ratgasdustmodel). We note that, as the models do not include radial drift or radially dependent grain growth, these gas--dust size differences are set only by optical depth. Here we compare \ratgasdustmodel\ to the observed gas--dust size difference (\ratgasdustobs), to identify which disks in our sample have undergone radial drift and radially dependent grain growth.  

Our analysis of \ratgasdust\ is based on the fact that our models reproduce the extent of the continuum emission. However, in the previous section we found a seemingly systematic offset in \rdust, namely $\rmmmodel > \rmmobs$, which might affect our interpretation of \ratgasdustobs versus \ratgasdustmodel. We therefore elect to propagate the effect of this offset into \ratgasdustmodel.
We fit the offset with a straight line ($\rmmmodel = a\times\rmmobs +b$, see Figure \ref{fig: gas and dust outer radii}) and use the best-fit values to scale \rmmmodel\ when calculating \ratgasdustmodel. We note that for the disks where $\rmmmodel > \rmmobs$ this will increase \ratgasdustmodel\ and enhance the perceived effect of optical depth on \ratgasdust. Therefore, for any disk showing $\ratgasdustobs > \ratgasdustmodel,$ we can confidently say that optical depth cannot explain their observed gas--dust size ratio.

\begin{figure}[htb]
    \centering
    \includegraphics[width=1.0\columnwidth]{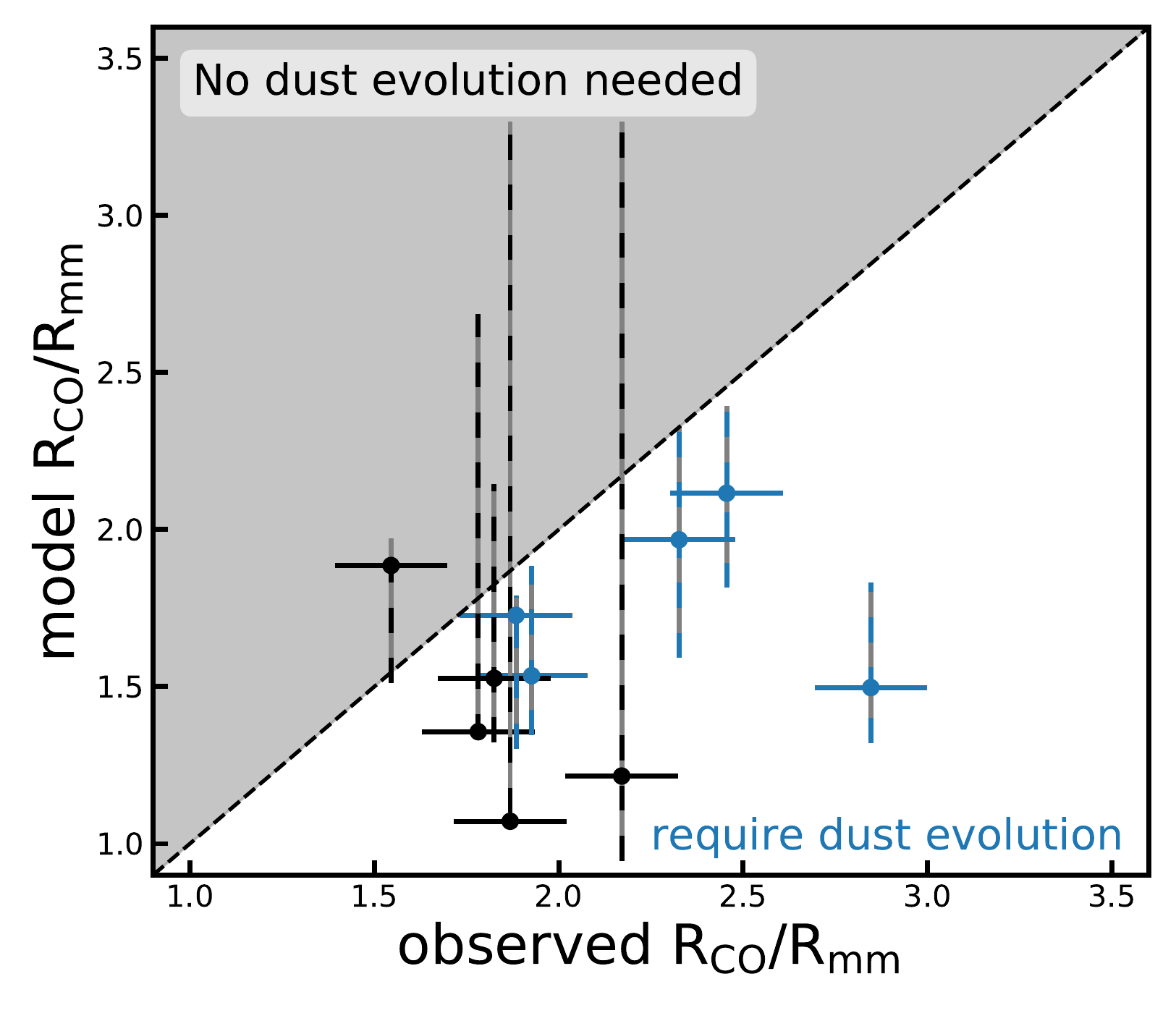}
    \caption{\label{fig: gas-dust size} Disk gas--dust size ratio comparison of the models with noise and the observations. Uncertainties of the model \ratgasdustmodel\ were computed using the 16$^{\rm th}$ and 84$^{\rm th}$ quantile of the \rgas\ distribution. Sources where \ratgasdustmodel\,$<$\,\ratgasdustobs\ are shown in blue. 
    To reproduce the observed \ratgasdustobs\ these sources require a combination of dust evolution and optical depth.
    }
\end{figure}

Figure \ref{fig: gas-dust size} shows \ratgasdust\ of the models and the observations. The gas--dust size ratios were calculated using the gas and dust outer radii shown in Figure \ref{fig: gas and dust outer radii}. The uncertainties on \ratgasdustobs\ were calculated using the uncertainties on \rcoobs\ and \rmmobs. The uncertainties on \ratgasdustmodel\ were calculated using the 16$^{\rm th}$ and 84$^{\rm th}$ quantile of the noisy \rcomodel\ distribution (see Appendix \ref{app: noisy rgas distributions}).

For five of the ten  disks in our sample (50\%)\ we find $\ratgasdustobs\,>\,\ratgasdustmodel$ even after the effects of noise are included, indicating this is a solid result considering the uncertainties on the measurements. These disks are Sz\,65, Sz\,71, J1600236, Sz\,98, and Sz\,129. For these disks, we need both dust evolution and optical depth effects to explain the observed \ratgasdustobs.

For the other five disks in the sample, Sz\,133, Sz\,100, MY Lup Sz\,84, and Sz\,68, the observed \ratgasdustobs\ lies within the uncertainty on \ratgasdustmodel. For these disks, it is possible that our model would reproduce the observed \ratgasdustobs\ if it were observed at similar sensitivity to the band 6 observations. We are however not able rule out that the \ratgasdustobs\ measured for these five sources are only due to optical depth effects. 

Among these latter five disks is Sz\,68, also known as HT\,Lup, which has been observed at high spatial resolution as part of the DSHARP program (\citealt{Andrews2018a}, see also \citealt{Kurtovic2018}). Sz\,68 is a multiple star system and the high-resolution observations were able to individually detect both the primary disk around Sz\,68 A and the disks around Sz\,68 B and C located at 25 and 434 AU in projected separation from Sz\,68 A \citep{Kurtovic2018}, respectively. In our observations with a resolution of $\sim$39 AU the disks around Sz\,68 A and B are not resolved separately. The observed \ratgasdustobs\ is therefore not likely to reflect the evolution of dust in this system.

It should be noted that of these five disks, three show large uncertainties towards large \ratgasdustmodel, which can be traced back to similarly large uncertainties on \rcomodel. 
\cite{Trapman2019} showed that a peak S/N $\geq 10$ on the moment-zero map of $^{12}$CO is required to measure \rgas\ to within 20\% (see, e.g., their Figure H.1). To compare this to our observations, the peak S/N in our sample varies from $\sim6$ (Sz 100) to $\sim12$ (Sz 133).
This suggests that for these three disks our comparison between \ratgasdustobs\ and \ratgasdustmodel\ is probably not reliable due to observational effects. 
Observations with a factor two higher sensitivity, equivalent to increasing the integration time from 1 to 4 minutes per source, would be sufficient to remove these observational effects. 

\section{Discussion}
\label{sec: discussion}

\subsection{Fast dust evolution candidates in Lupus}
\label{sec: analyzing remaining disks}

\begin{figure*}[htbp]
     \centering
    \includegraphics[width=0.8\textwidth]{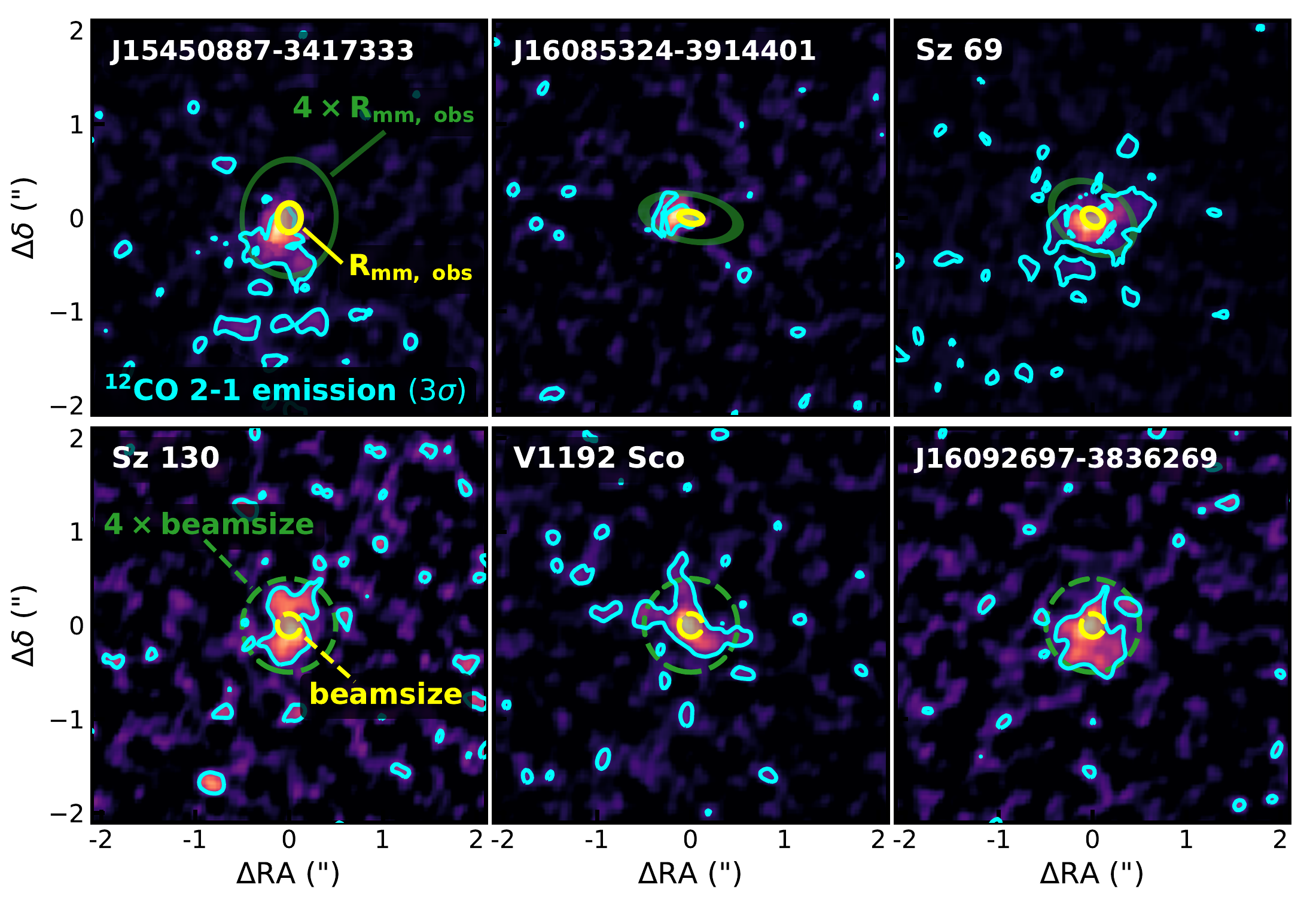}
    \caption{\label{fig: drift candidates} \CO moment-zero maps of the six sources from the ``low S/N sample'' (see Section \ref{sec: analyzing remaining disks}) 
    where the S/N = 3 contour of their $^{12}$CO emission, shown in cyan, reaches beyond  $4\times\rdust$. Using this contour as a proxy for \rgas, these disks likely have $\ratgasdust \geq 4$ and are therefore clear candidates for having undergone dust evolution (cf. \citealt{Trapman2019}). 
    The top three disks have resolved continuum emission and their $\rmmobs$ is shown by the yellow ellipse. For these sources, Keplerian masking was applied to the $^{12}$CO $J = 2\,-\,1$ emission (see Appendix \ref{app: keplerian masking}). 
    The bottom three disks have unresolved continuum emission. The dashed yellow circle shows the size of the beam $(0\farcs25)$ as an upper limit to the dust disk size. Similarly, the dashed green circle shows four times the beam size. }
\end{figure*}

In the present study, we look at the gas--dust size differences for a sample of 10 of the 48 disks in Lupus where $^{12}$CO $J = 2\,-\,1$ is detected. Our sample makes up $\sim15\%$ of the Lupus disk population and is biased towards the most massive ($\geq 10 \mathrm{M}_{\oplus}$) dust disks (see Figure \ref{fig: sample properties}). Notably, the $^{12}$CO detections are not similarly biased, with $^{12}$CO $J = 2\,-\,1$ being detected for some of the faintest continuum sources. 

Here we examine the 38 disks in Lupus where $^{12}$CO $J = 2\,-\,1$ was detected, but which did not meet our selection criteria; we refer to these as the ``low-S/N
sample''. As discussed in Section \ref{sec: sample selection}, most of these sources were excluded from our analysis because the S/N of the $^{12}$CO $J = 2\,-\,1$ was too low to measure \rgas. Analyzing them in the same manner as the ``high S/N sample'' is not possible with the current observations.
The disks in the high S/N sample show that \rgas\ approximately coincides with a contour showing S/N = 3 $^{12}$CO $J = 2\,-\,1$ emission. For the disks in the low S/N sample we can therefore use the S/N = 3 contour of $^{12}$CO $J = 2\,-\,1$ as a proxy for \rgas, giving us some idea of their gas disk sizes. We note that this approximation is likely to underestimate \rgas, as observing these sources at a sensitivity matching the high S/N sample would move the S/N = 3 contour outward. 

With this proxy for \rgas, we can investigate whether the disks in the low S/N sample show signs of dust evolution. Using \texttt{DALI,} \cite{Trapman2019} compared the \ratgasdustmodel\ of a series of models with and without dust evolution.
These latter authors found that $\ratgasdust \geq 4$ is a clear sign of dust evolution, giving us a clear criterion with which we can identify signs of dust evolution.
If a disk in the low S/N sample has a S/N = 3 contour for its $^{12}$CO emission that reaches beyond $4\times\rdust$, it is likely that this disk would have $\ratgasdust \geq 4$ if observed at higher sensitivity. We therefore identify it as a disk showing clear signs of having undergone dust evolution, where it would be difficult to explain the observations using only optical depth and without any radial drift or radially dependent grain growth.

As shown in Figure \ref{fig: drift candidates} we identify six disks where the S/N = 3 contour of their $^{12}$CO emission reaches beyond $4\times\rdust$
Of these, three have marginally resolved 1.3 mm continuum emission, namely J15450887\,-\,3417333, J16085324\,-\,3914401, and Sz~69. 
Although \rgas\ could not be measured for these sources, it is difficult to explain the difference between the extent of the CO and the extent of the continuum without including dust evolution. 
It should be noted that the $^{12}$CO channel maps of J15450887 show some cloud emission. Most of this emission is removed by the Keplerian mask but some of it could still be present in the moment-zero map.

The other three disks remain unresolved in the continuum at a resolution of $0\farcs25$. As an upper limit for the dust disk size we use $0\farcs125$ which is approximately the radius of the beam. These three disks, Sz~130, V1192~Sco, and J16092697\,-\,3836269, show significant CO emission outside $4\times\frac{1}{2}\times{\rm beam width} = 4\times0\farcs125$. Taking into account that the dust disk of these sources is unresolved, it is very likely that these disks have undergone substantial dust evolution. Noteworthy here is the inclusion of V1192~Sco, which has the faintest detected continuum flux of the disks in Lupus. Several studies have shown that there exists a relation between the millimeter luminosity and the dust disk size (see, e.g., \citealt{Tazzari2017,Tripathi2017,Andrews2018a}).
If we extrapolate this relation down to the observed millimeter flux of V1192~Sco, we find a dust disk size of 4-8 AU $(0\farcs025-0\farcs05)$. With $^{12}$CO $J = 2\,-\,1$ emission extending out up to $0\farcs75$, V1192~Sco 
would seemingly have a gas--dust size difference of $\ratgasdustobs~15-30$, which would make it one of the most extreme cases of grain growth and radial drift. Deeper observations of $^{12}$CO and higher resolution observations of the continuum are required to confirm this.
We should note here that these disks could have faint, extended continuum emission that was undetected in our current observations.
However, given the sensitivity of our current observations this faint emission can at most increase \rdust\ by a factor of two for our faintest source, V1192~Sco.

\subsubsection{Are compact dust disks the result of runaway radial drift?}
\label{sec: radial drift efficiency}

The discovery of six disks that likely have $\ratgasdustobs \geq 4$ highlights an interesting property of the ten disks analyzed in detail in here, namely that they all have relatively low gas--dust size differences. The disks in our sample have $\langle \ratgasdustobs\rangle_{\rm obs} = 2.06 \pm 0.37$ where the second number specifies the standard deviation of the sample. As discussed in Section \ref{sec: sample selection}, these disks represent the massive end of the Lupus disk population (see Figure \ref{fig: sample properties}). 
In contrast, there are at least six disks with low dust masses that have $\ratgasdustobs \geq 4$. These are all disks with compact continuum emission ($\rdust \leq 24$\,AU).
We would expect more massive disks to have a higher \ratgasdust\ (see, e.g., \citealt{Trapman2019}). As a result of their higher disk mass, these disks have a greater total CO content which results in a larger observed gas outer radius (\rgas). In addition, these disks have a higher dust mass, resulting in more efficient grain growth and inward radial drift.

The low \ratgasdustobs\ in our sample could be linked to the rings and gaps observed in large disks (see, e.g., \citealt{Andrews2018b,Huang2018}). These rings are the result of a local pressure maximum that acts as a dust trap for the millimeter-sized grains (see, e.g., \citealt{Whipple1972, KlahrHenning1997,KretkeLin2007,Pinilla2012}). These dust traps stop the millimeter-sized grains from drifting inward, effectively halting the radial dust evolution. As a result, the disks stay both large and bright in terms of the millimeter continuum emission. 
This hypothesis is also in line with recent results from a high-resolution ALMA survey in Taurus \citep{Long2018,Long2019}. Of the 32 disks observed at $\sim$16 AU resolution, all disks with a dust outer radius of at least 55 AU show detectable substructures \citep{Long2019}, whereas all disks without substructures are found to be small. \cite{Long2019} hypothesize that fast radial drift could be the cause of this dichotomy.  

The disks in our sample are massive ($\mdust = 10-200\ \mathrm{M}_{\oplus}$) and should have been capable of forming gap-opening planets in the outer part of the disk. Using high-resolution observations of GW Lup (Sz\,71), \cite{Zhang2018} inferred the presence of a $\sim10$ M$_{\oplus}$ planet at 74 AU based on a gap in the continuum emission. The six disks of our sample with $\ratgasdustobs \geq 4$ have a much lower dust mass ($\mdust = 0.4-10\ \mathrm{M}_{\oplus}$) and could have been unable to form a gap-opening planet in their outer disk. Without these gaps acting as dust traps, radial drift would be unimpeded, leading to a compact disk of millimeter-sized grains and a large gas--dust size difference. 

Alternatively, the disks in our sample could have a higher level of CO underabundance, and therefore a lower total CO content, which would explain their low \ratgasdustobs\ (see, e.g., Section \ref{sec: total co content}; \citealt{Trapman2019}). Being both large and massive, the disks in our sample are expected to be cold, leading to a larger fraction of CO being frozen out. A lower total CO content leads to a smaller \rgas\ and a lower \ratgasdust. 

Being compact and less massive, the six disks with $\ratgasdustobs \geq 4$ are expected to be warmer. In these disks it would be harder to keep CO frozen out on the grains, lowering the effectiveness of the processes suggested to be responsible for the observed CO underabundance (see Section \ref{sec: total co content} and references therein). The lack of a CO underabundance would result in larger \rgas\ and a higher \ratgasdust. However, \citealt{miotello2017} showed that Sz\,90, J15450887\,-\,3417333, Sz\,69, and Sz\,130 have $\Delta_{\rm gd} \geq 10$, indicating that they have a similar level of CO underabundance to the ten disks in our sample.

\section{Conclusions}
\label{sec: conclusions}

In the present study, the observed gas and dust size dichotomy in protoplanetary disks was studied in order to investigate the occurrence of common radial drift and radially dependent grain growth across the Lupus disk population. The gas structure of a sample of ten disks in the Lupus star-forming regions was modeled in detail using the thermochemical code DALI \citep{Bruderer2012,Bruderer2013}, incorporating the effects of CO isotope-selective processes \citep{Miotello2014}. Surface density structures were based on modeling of the continuum emission by \cite{Tazzari2017}. The total CO content of the models was fitted using integrated $^{13}$CO fluxes to account for either gas depletion or CO underabundance. 
Noise was added to the synthetic $^{12}$CO emission maps and gas and dust outer radii were measured from synthetic $^{12}$CO and 1.3 millimeter emission maps using the same steps used to measure these quantities from the observations. 
From comparisons of our model gas and dust outer radii to the observations, we draw the following conclusions:

\begin{itemize}
    \item For five disks (Sz\,98, Sz\,71, J16000236\,-\,4222115, Sz\,129 and Sz\,65) we find $\ratgasdustobs\,>\,\ratgasdustmodel$. For these disks we need both dust evolution and optical depth effects to explain the observed gas--dust size difference.
    \item For five disks (Sz\,133, MY Lup, Sz\,68, Sz\,84 and Sz\,100), the observed \ratgasdustobs\ lies within the uncertainties on \ratgasdustmodel\ due to noise. For these disks the observed gas--dust size difference can be explained using optical line effects only.
    \item We identify six disks without a measured \rgas\ that show significant (S/N $\geq3$) $^{12}$CO $J = 2\,-\,1$ emission beyond $4\times\rdust$. These disks likely have $\ratgasdust \gg 4,$ which would be difficult to explain without substantial dust evolution.
    \item The wide range of noisy \rcomodel\ measured for the two smallest disks in our sample show that  care should be taken when measuring the gas disks size of faint compact disks using a curve of growth method.
\end{itemize}

Our analysis shows that most of the disks in our sample, which represent the bright end of the Lupus disk population, are consistent with radial drift and grain growth. Furthermore, we also find six faint disks with $^{12}$CO emission beyond four times their dust disk size, suggesting that radial drift is a common feature among bright and faint disks. 
For both cases, our analysis is limited by the sensitivity of current disk surveys. 
More sensitive disk surveys that integrate 5\,-\,10 minutes per source are required to obtain a complete picture of radial drift and grain growth in ``typical disks'' in young star-forming regions.

\begin{acknowledgements}
We would like to thank the referee for constructive and detailed comments that greatly improved the presentation of the paper.
We would also like to thank A. Bosman, L. Testi and M. Tazzari for the useful discussions. 
LT and MRH are supported by NWO grant 614.001.352. M.A. acknowledges support from NSF AST-1518332, NASA NNX15AC89G and NNX15AD95G/NEXSS. This work benefited from NASA's Nexus for Exoplanet System Science (NExSS) research coordination network sponsored by NASA's Science Mission Directorate. Astrochemistry in Leiden is supported by the European Union A-ERC grant 291141 CHEMPLAN, by the Netherlands Research School for Astronomy (NOVA), and by a Royal Netherlands Academy of Arts and Sciences (KNAW) professor prize. SF and CFM are supported by ESO fellowships. JPW is supported NASA grant NNX15AC92G.
This paper makes use of the following ALMA data: 2013.1.00220.S, 2015.1.00222.S, 2013.1.00226.S, 2013.00694. ALMA is a partnership of ESO (representing its member states), NSF (USA) and NINS (Japan), together with NRC (Canada), NSC and ASIAA (Taiwan), and KASI (Republic of Korea), in cooperation with the Republic of Chile. The Joint ALMA Observatory is operated by
ESO, AUI/NRAO and NAOJ. All figures were generated with the \texttt{PYTHON}-based package \texttt{MATPLOTLIB} \citep{Hunter2007}.

\end{acknowledgements}

\bibliographystyle{aa} 
\bibliography{references}


\begin{appendix}

\section{Keplerian masking}
\label{app: keplerian masking}

Analysis of the gas emission of protoplanetary disks is often performed using the moment-zero map, which is obtained by integrating the observed spectral cube along the velocity axis. The velocity range for the integration is set by the maximum velocity offset (positive and negative) relative to the source velocity where emission of the disk is still observed. This leads to broad velocity integration range, whereas in the outer regions of the disk line emission is coming from a much more narrower velocity range. The S/N in these regions can therefore be improved if the integration is limited to only those channels containing emission. 

This method of improving the S/N of moment-zero maps has already been used (e.g., \citealt{Salinas2016,Carney2017,Bergner2018,Loomis2018b}). \cite{Yen2016} developed a similar method whereby spectra are aligned at different positions in the disk by shifting them by the projected Keplerian velocities at their positions and are then stacked. Their method produces an aligned spectrum with increased S/N, well suited for detecting emission not seen in individual channels (e.g., \citealt{Yen2018}).

The masking method works for observed line emission from a source with a known, ordered velocity pattern. In the case of protoplanetary disks, this is the Keplerian rotation around the central star. Using prior knowledge of this rotation pattern, voxels of the spectral cube can be selectively included in the analysis of the data, for example when making a moment-zero map, based on the criterion

\begin{equation}
\label{eq: criterion example}
|V_{\rm kep} (\alpha,\delta) -V_{\rm voxel}| \leq \frac{1}{2} \cdot \mathrm{channel\ width}.
\end{equation}

\noindent Here, $V_{\rm kep}$ is the Keplerian velocity at the coordinates (RA, Dec) = $(\alpha,\delta)$ and $V_{\rm voxel}$ is the velocity coordinate of the voxel.

\subsection{Implementation}
\label{sec: implementation}

\subsubsection{Calculate the projected Keplerian velocity pattern}
\label{sec: calculate velocities}

In order to calculate the Keplerian velocities at a given point $p$ with coordinates $(\alpha,\delta)$ in the observed image, the coordinates of $p$ first have to be projected onto the 2D local frame of the source. For a source with an observed position angle $PA$ and inclination $i$ at a distance $d$ this coordinate transformation is given by

\begin{align}
\label{eq: projection}
    x &= \frac{1}{\cos i} \left( \left(\alpha - \alpha_0\right) \cos PA + \left(\delta -\delta_0\right) \sin PA \right)\cdot d\\
    y &= \left( -\left(\alpha - \alpha_0\right) \sin PA + \left(\delta -\delta_0\right) \cos PA \right)\cdot d. 
\end{align}
Here $\alpha,\delta$ are the right ascension and declination of the point $p$ and ($\alpha_0,\delta_0$) is the location of the center of the source.

Using the local coordinates $x,y$ the Keplerian velocity at $p$ can be calculated using
\begin{equation}
\label{eq: keplerian velocity}
V_{\rm kep} = \sqrt{\frac{G M_*}{r}};\ \ r =\sqrt{x^2 + y^2}.
\end{equation}
Here $G$ is the gravitational constant, $M_*$ is the stellar mass, and $r$ is the deprojected radial distance from the star.

To convert the Keplerian velocity back to the velocity at which the emission will be observed, it has to be projected along the line of sight:

\begin{equation}
\label{eq: projected velocity}
V_{\rm proj} = - V_{\rm kep} \sin i \frac{y}{r} + V_{\rm sys},
\end{equation}
where $V_{\rm sys}$ is the systematic velocity of the source. 

\subsubsection{Selecting voxels containing emission}
\label{sec: selecting voxels}

For a given point $p = (\alpha,\delta)$ equation \eqref{eq: projected velocity} gives the expected velocity of the emission. Based on this information a Keplerian mask can be created by selecting a voxel $nml$ for the Keplerian mask if
\begin{equation}
\label{eq: criterion keplerian mask}
\left| V_{\rm proj} (\alpha_n,\delta_m | M_*, PA, i, \alpha_0, \delta_0, V_{\rm sys}) -V_l \right| \leq \frac{1}{2} \Delta V_{\rm width} + \Delta V_{int}.
\end{equation}
Here $\alpha_n, \delta_m, V_l$ are the coordinates of the voxel and $\Delta V_{\rm width}$ is the channel width. Further,  $\Delta V_{\rm int}$ is introduced in Eq. \eqref{eq: criterion keplerian mask} to compensate for the fact that using a Keplerian rotation profile is a simplification that is  only valid if the disk is geometrically thin and the rotation is purely Keplerian. In reality, the line emission is more likely to originate from layers higher in the (often flared) disk (e.g., \citealt{Dutrey2014}). As a result a single pixel, representing a single line of sight through the disk, contains contributions from different vertical layers at project velocities that are offset from the Keplerian rotation velocity of the midplane. 

Here $\Delta V_{\rm int}$ is left as a free parameter with no radial dependence in order to make no assumptions on the vertical structure of the disk. However, such a dependence could be introduced. For example \cite{Yen2016} use the empirically fitted description for $\Delta V_{\rm int}$ from \cite{Pietu2007}.

\subsubsection{Convolving the mask}
\label{sec: convolving the mask}

After the mask is set up, it is convolved with the beam to include the effects of resolution on the channel maps.
We note that this effect is only relevant if the smearing by the beam is much larger than the width of the channel. 

\subsubsection{Clipping the mask}
\label{sec: clipping the mask}

In order to ensure flux conservation in the masked region, a final step has to be made. After the convolution in the previous step, the pixels of the mask now have weights $\leq 1$. The mask is therefore converted back into a boxcar function according to 

\begin{equation}
\label{eq: clipped mask}
M_{\rm clipped}(\alpha_n, \delta_m,V_l) = \begin{cases}
                            1 & M(\alpha_n, \delta_m,V_l) \leq \text{cutoff} \\
                            0 & \text{else}, \\
                            \end{cases}
\end{equation}
where $M$ is the mask after step 3 and the cutoff is set to $0.05$ of the peak value.

\subsection{Making moment-zero maps and calculating noise}
\label{sec: making mom0 maps}

After the mask has been produced following the steps mentioned above, it can be applied to the data. The moment-zero map can be calculated following

\begin{equation}
\label{eq: moment 0}
\text{Mom 0}(\alpha_n,\delta_m) = \sum_{l=0}^L M_{\rm clipped}(\alpha_n, \delta_m,V_l)\times I(\alpha_n, \delta_m,V_l),
\end{equation}
where $I(\alpha_n, \delta_m,V_l)$ is the observed spectral cube and $V_0 $ and $V_L$ are the minimum and maximum velocities of the spectral cube respectively. 

In a similar manner, a map of the expected noise levels in the moment-zero map can be calculated. As a result of the masking, individual pixels will have different noise characteristics depending on how many nonzero voxels in the mask are summed over (cf. Eq. \eqref{eq: moment 0}. Using the fact that the noise between individual channels is independent, a 2D noise map can be created using

\begin{equation}
\label{eq: noise}
N(\alpha_n,\delta_m) = \text{RMS}\times \sqrt{\sum_{l=0}^L \left(M_{\rm clipped}(\alpha_n, \delta_m,V_l) \right)^2 },
\end{equation}
where RMS is the root mean square noise taken from an empty channel.

As a result of the term in the square root, the S/N defined as $\text{S/N} \equiv \text{Mom 0}/N$ can be increased  by not clipping the mask (Section \ref{sec: clipping the mask}), but this comes at the cost of a reduced total flux in the moment-zero map. This difference can be understood by the fact that the convolved mask provides lower weights ($w_{nml}$) for voxels that are expected to contain very little flux with respect to the noise in that pixel. In the noise, these voxels are almost excluded due to the term $w_{nml}^2$ in Eq \eqref{eq: noise}, resulting in a lower noise for that pixel. In the moment-zero map however, the flux in these voxels is also scaled down by a factor $w_{nml}$. As $w_{nml} \leq 1$, flux is no longer conserved in this case.

\subsection{Caveats}
\label{sec: caveats}

In step one of making the mask (Section \ref{sec: calculate velocities}), a simplification is made in that a single velocity can be assigned to a pixel, i.e., that the velocity gradient over the length of pixel is small. At the center this simplification breaks down. To circumvent this problem the center pixel can be included in all channels.

We note that as a consequence the S/N in the center part decreases to the pre-masked values. For most sources this is not a significant problem, as the S/N is usually highest at the center where the source is brightest. 

\subsection{The Keplerian mask parameters of our sample}
\label{sec: masking in our work}

Here we outline the Keplerian mask used in this work. As shown in Eq. \eqref{eq: criterion keplerian mask} the Keplerian mask is described by seven parameters: the stellar mass ($M_*$), the orientation of the disk ($PA,i$), the three coordinate centroids ($\alpha_0,\delta_0, V_{\rm sys}$), and the free parameter $\Delta V_{\rm int}$. For the stellar masses we use the observations and methods presented in \cite{alcala2014,alcala2017}, but rescaled to the new \emph{Gaia} DR2 distances (\citealt{GaiaDR2_2018}, see also Appendix A of \citealt{Manara2018}). The position angle, inclination, and centroid were taken from the observations of the millimeter continuum (cf. Tables 1 and 2 in \citealt{Tazzari2017}). The final two parameters, $V_{\rm sys}$ and $\Delta V_{\rm int}$, were obtained by varying them to maximize the total S/N in the moment-zero map. The mask parameters of the ten sources in our sample plus the eight sources with resolved continuum emission described in Section \ref{sec: analyzing remaining disks} are presented in Table \ref{tab: mask parameters}

\begin{table*}[htb]
\centering
\caption{\label{tab: mask parameters} Keplerian masks}
\begin{tabular*}{0.71\textwidth}{l|ccccc|cc}
\hline\hline
Name & M$_*$ & PA & $i$ &  $\alpha_0$ & $\delta_0$ & $V_{\rm sys}$ & $\Delta V_{\rm int}$ \\
 & (M$_{\odot}$) &  (deg)  & (deg) & (J2000) & (J2000) & (km s$^{-1}$) & (km s$^{-1}$)\\
\hline
Sz~133    & 0.63 & 126 & 79 & 16:03:29.37 & -41:40:02.14 & 4.22 & 0.83 \\ 
Sz~98     & 0.67 & 112 & 47 & 16:08:22.48 & -39:04:46.81 & 2.81 & 0.62 \\
MY~Lup    & 1.09 &  59 & 73 & 16:00:44.50 & -41:55:31.27 & 4.5  & 1.11 \\
Sz~71     & 0.41 &  38 & 41 & 15:46:44.71 & -34:30:36.05 & 3.2  & 0.7 \\
J16000236 & 0.23 & 160 & 66 & 16:00:02.34 & -42:22:14.99 & 4.0  & 0.65 \\
Sz~129    & 0.78 & 155 & 32 & 15:59:16.45 & -41:57:10.66 & 4.2  & 0.36 \\
Sz~68     & 2.13 & 176 & 33 & 15:45:12.84 & -34:17:30.98 & 4.9  & 0.6 \\
Sz~100    & 0.14 &  60 & 45 & 16:08:25.74 & -39:06:01.63 & 1.9  & 1.2 \\
Sz~65     & 0.7  & 109 & 61 & 15:39:27.75 & -34:46:17.56 & 4.4  & 0.89 \\
Sz~84     & 0.17 & 167 & 74 & 15:58:02.50 & -37:36:03.08 & 5.2  & 1.6 \\
\hline
J15450887 & 0.14 &   2 & 36 & 15:45:08.85 & -34:17:33.81 & 4.5  & 0.9 \\
J16085324 & 0.02 & 100 & 61 & 16:08:53.22 & -39:14:40.53 & 3.0  & 1.0 \\
Sz~69     & 0.2  & 124 & 44 & 15:45:17.39 & -34:18:28.66 & 5.3  & 0.8 \\
Sz~83     & 0.67 & 164 &  3 & 15:56:42.29 & -37:49:15.82 & 4.23 & 0.2 \\
Sz~90     & 0.78 & 123 & 61 & 16:07:10.05 & -39:11:03.64 & 3.2  & 0.83 \\
Sz~73     & 0.78 &  95 & 50 & 15:47:56.92 & -35:14:35.15 & 4.1  & 0.93 \\
Sz~114    & 0.19 & 149 & 16 & 16:09:01.83 & -39:05:12.79 & 5.0  & 0.28 \\
J16124373 & 0.45 &  23 & 44 & 16:12:43.73 & -38:15:03.40 & 4.0  & 0.75 \\
J16102955 & 0.2  & 119 & 67 & 16:10:29.53 & -39:22:14.83 & 3.5  & 1.2 \\
\end{tabular*}
\captionsetup{width=.6\textwidth}
\caption*{\footnotesize{ }}
\end{table*}

\newpage

\section{Influence of dust settling and flaring}
\label{app: flaring and dust settling}

In our models we have kept the disk vertical structure fixed. In addition we assume a single height and mass fraction of large grains (cf. Section \ref{sec: vertical structure} and \ref{sec: dust settling}).
The vertical structure and the distribution of the large grains set both the temperature structure and the chemistry. Varying them could therefore change the $^{13}$CO 3\,-\,2 flux used to determine the total CO content and the shape of $^{12}$CO intensity profile from which we measure \rgas. We examine the effect of varying the  vertical structure and the distribution of the large grains on \rgas\ for two disks in our sample that represent two completely different physical structures: Sz 68 (compact, strong CO flux, $^{13}$CO optically thick)  and Sz 98 (large, weak CO flux, $^{13}$CO optically thin).

\begin{figure}[!htb]
\centering
\includegraphics[width=\columnwidth]{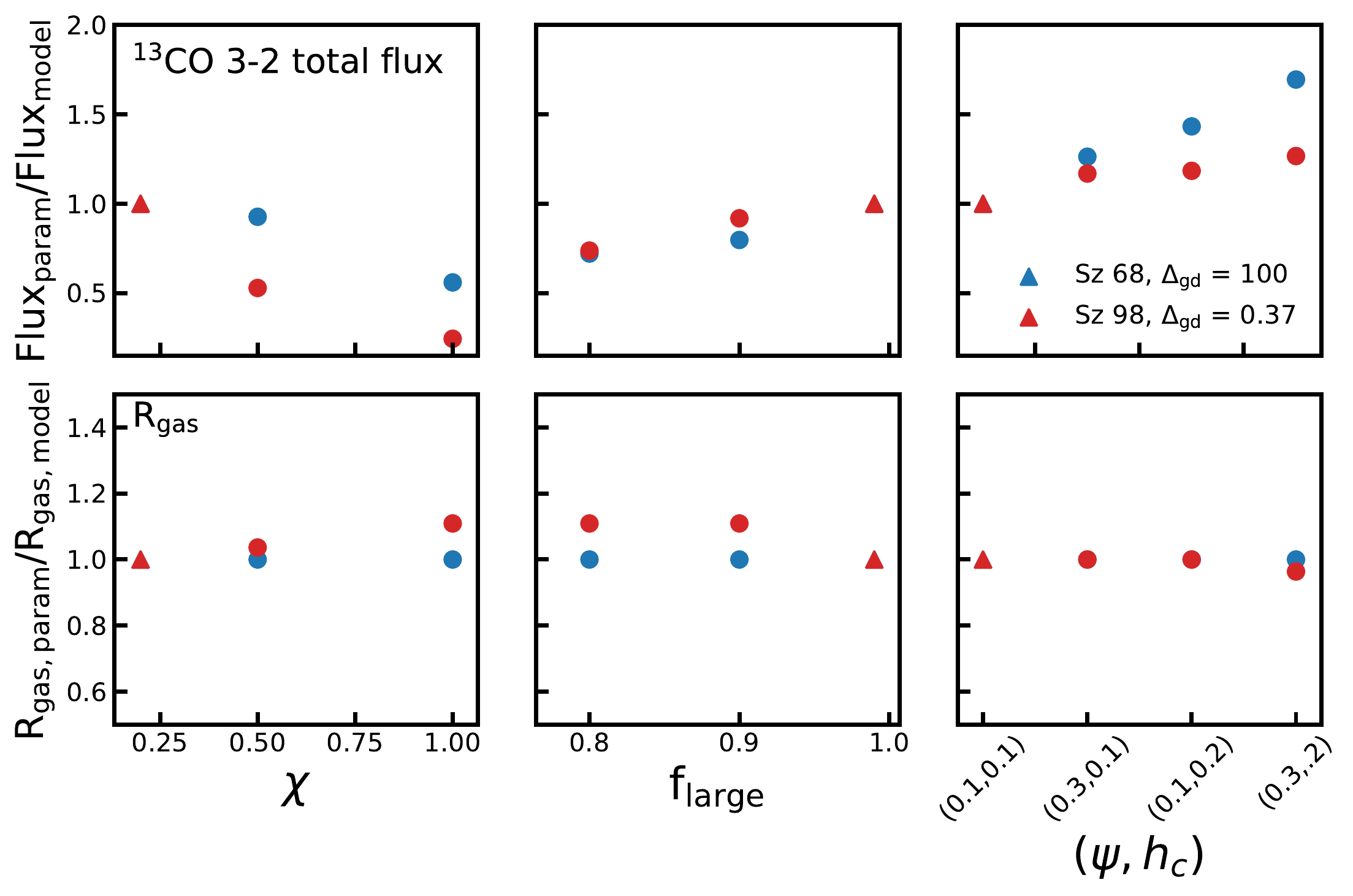}
\caption{\label{fig: effect other parameters} Effects of vertical structure and the large grains. \textbf{Top panels:} Integrated $^{13}$CO flux as function of large grain settling ($\chi$), fraction of large grains ($f_{\rm large}$) and disk vertical structure ($h = h_c (R/R_c)^{\psi}$). The models for Sz 98 and Sz 68 are shown in red and blue respectively. Triangle markers denote the value used in the rest of this work (cf. Sections \ref{sec: vertical structure} and \ref{sec: dust settling}). \textbf{Bottom panels:} As top panels, but showing the variations of the gas outer radius \rgas. }
\end{figure}

The results are shown in Figure \ref{fig: effect other parameters}. The top three panels show the effect on the $^{13}$CO integrated flux used to determine the total CO content of the disk (cf Section \ref{sec: total co content}). Increasing the vertical extent of the large grains ($\chi$) lowers the $^{13}$CO flux by up to $\sim$40\% for the optically thin Sz 98 and up to $\sim$80\% for the optically thick Sz 68. This is likely due to the dust becoming optically thick at millimeter wavelengths higher up in the disk. Decreasing the mass fraction of large grains ($f_{\rm large}$) also lowers the $^{13}$CO flux, up to $\sim$25\% for both disks. 
Changing either $\chi$ or $f_{\rm large}$ would require increasing the total CO content to reproduce the $^{13}$CO flux, which would increase \rgas. 

Changing the vertical structure of the disk by either increasing the amount of flaring or increasing the scale height of the disk will lead by to an increase in the observed $^{13}$CO integrated flux. Both parameters directly affect the amount of stellar light intercepted by the disk, leading to a higher temperature. The $^{13}$CO flux is increased by up to $\sim$20\% for Sz 68 and up to $\sim$65\% for Sz 98. Increasing either the flaring or the scale height would mean a lower total CO content is needed to match the observed $^{13}$CO flux, leading to a smaller \rgas\ (cf. \citealt{Trapman2019}).

The bottom three panels of Figure \ref{fig: effect other parameters} show how \rgas\ is affected by changes in $f_{\rm large},\,\chi,\,h_c,\,\Psi$. For Sz~98 the gas outer radius increases by less than 10\% if either $f_{\rm large}$ or $\chi$ is changed and the \rgas\ of Sz~68 is not affected at all. Changing the vertical structure of either disk changes the derived R$_{\rm gas}$ by less than 5\%. The vertical structure does affect the $^{12}$CO intensity profile, but the relative changes remain nearly constant over the extent of the disk. As a result the curve of growth and the inferred R$_{\rm gas}$ remain unaffected. 

\section{Radial 890~$\mu$m continuum profiles}
\label{app: continuum profiles}

\begin{figure*}
\centering
\begin{subfigure}{0.48\textwidth}
\includegraphics[width=\columnwidth]{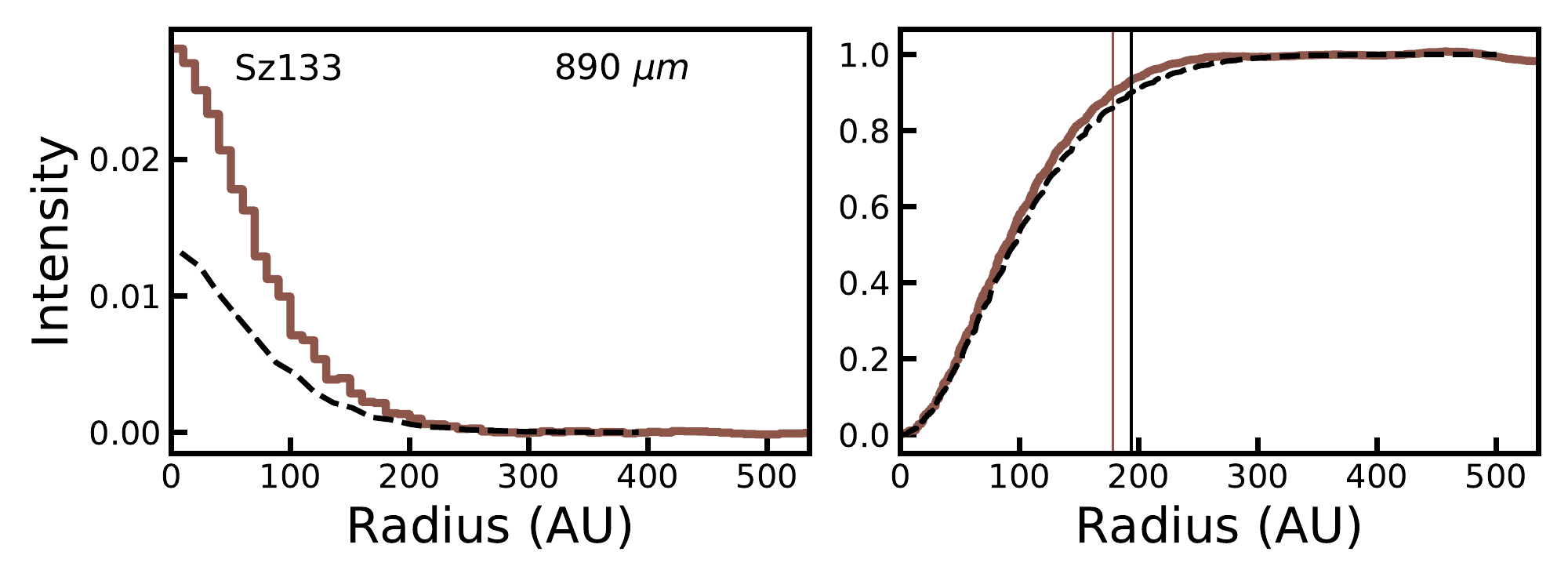}
\end{subfigure}
\begin{subfigure}{0.48\textwidth}
\includegraphics[width=\columnwidth]{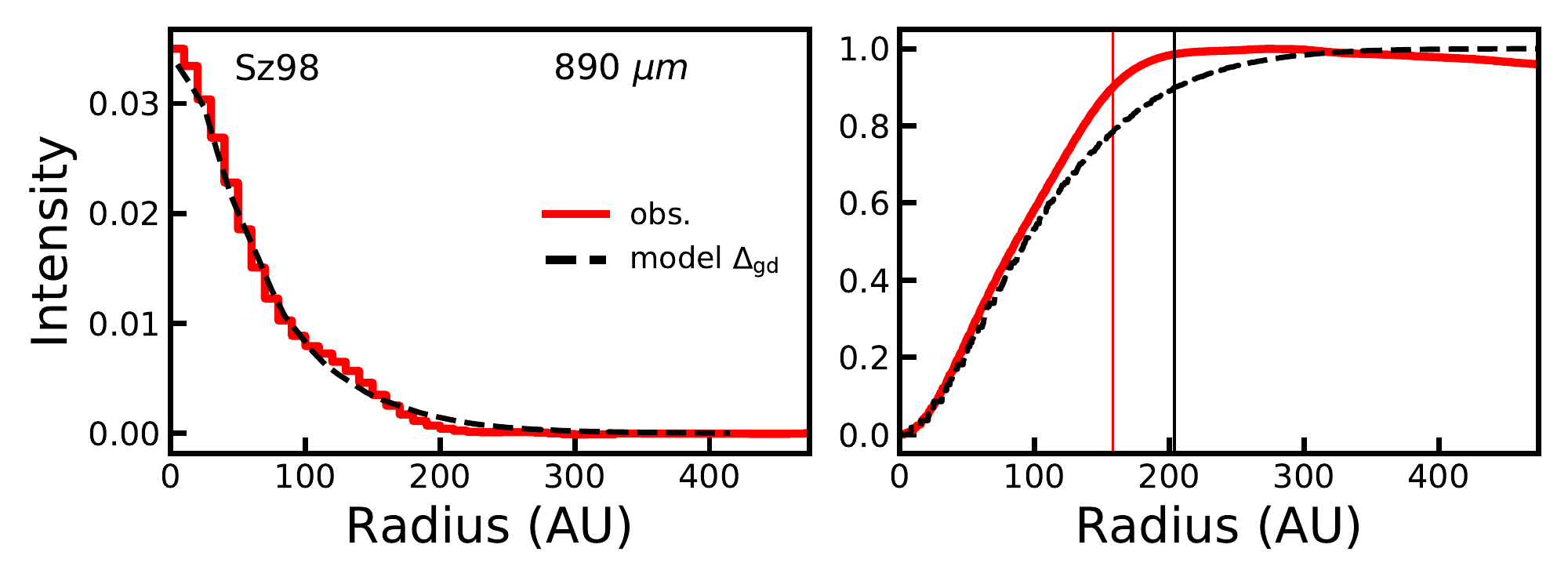}
\end{subfigure}

\begin{subfigure}{0.48\textwidth}
\includegraphics[width=\columnwidth]{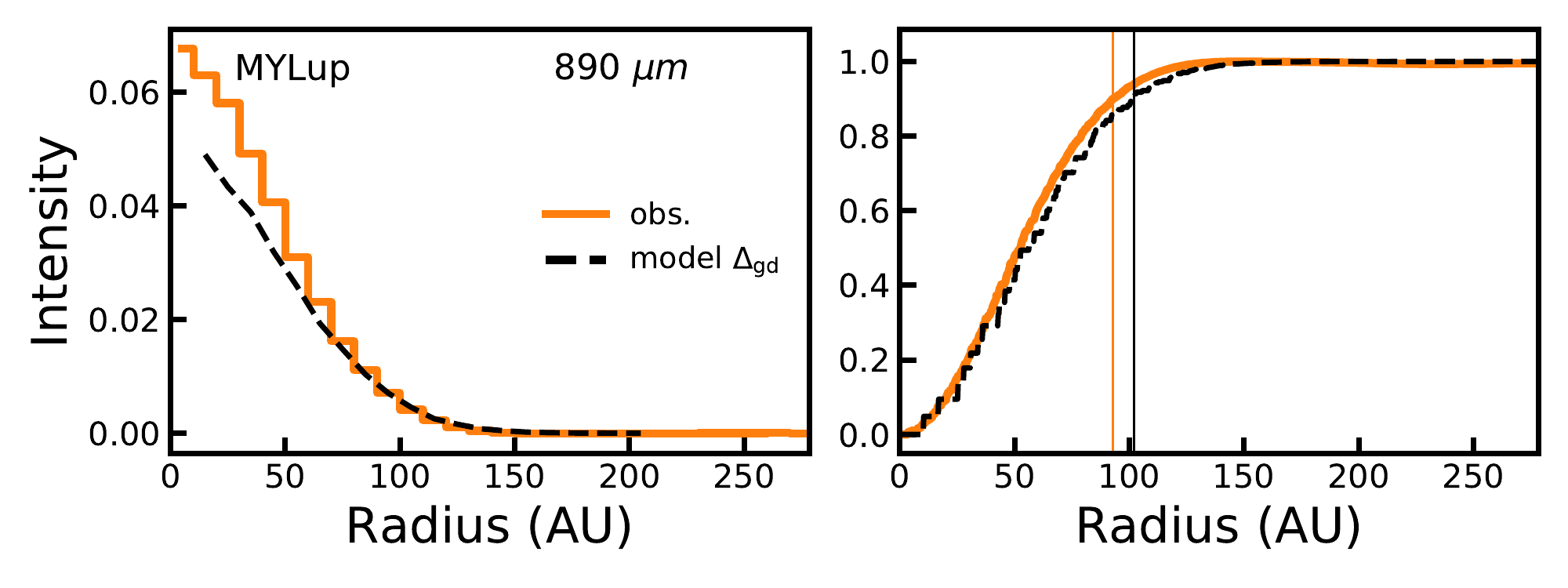}
\end{subfigure}
\begin{subfigure}{0.48\textwidth}
\includegraphics[width=\columnwidth]{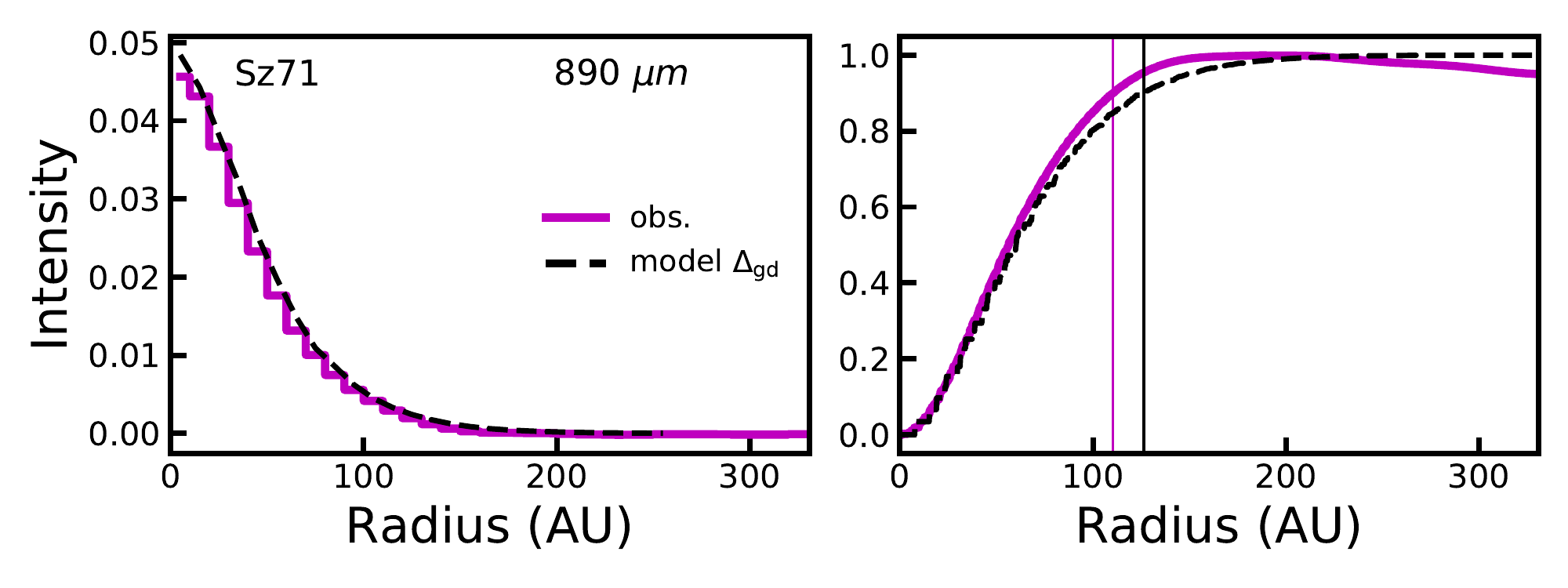}
\end{subfigure}

\begin{subfigure}{0.48\textwidth}
\includegraphics[width=\columnwidth]{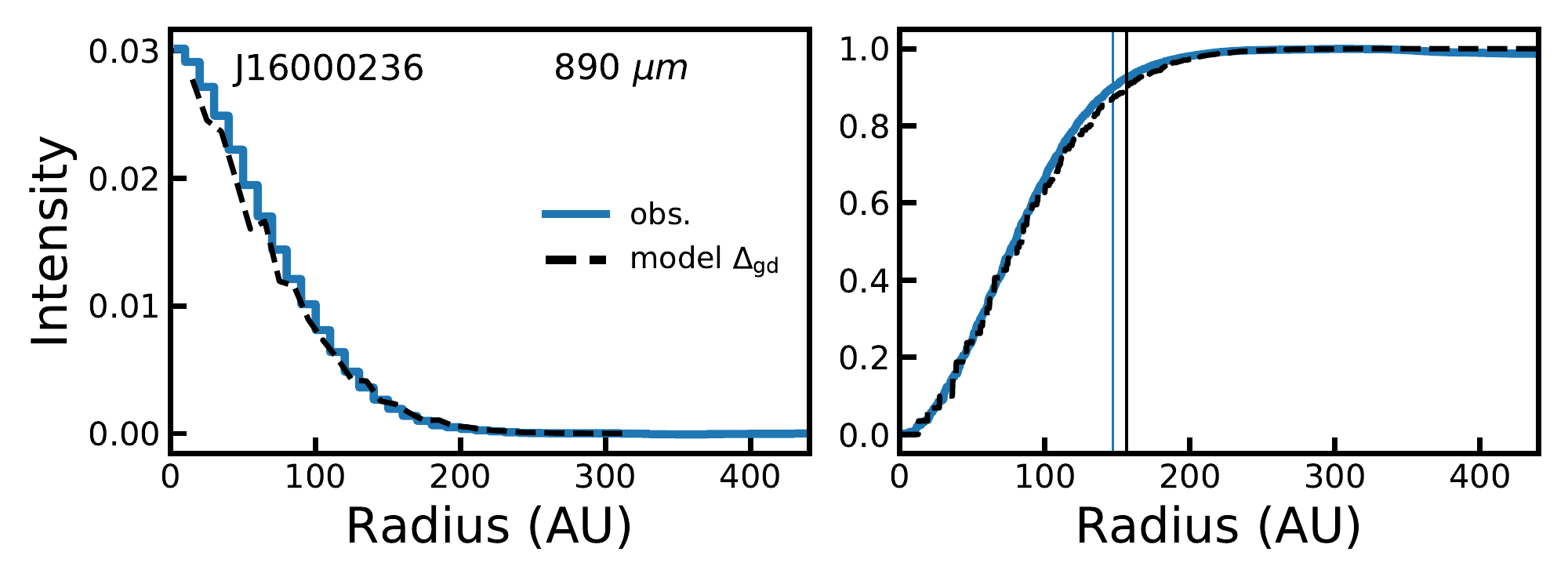}
\end{subfigure}
\begin{subfigure}{0.48\textwidth}
\includegraphics[width=\columnwidth]{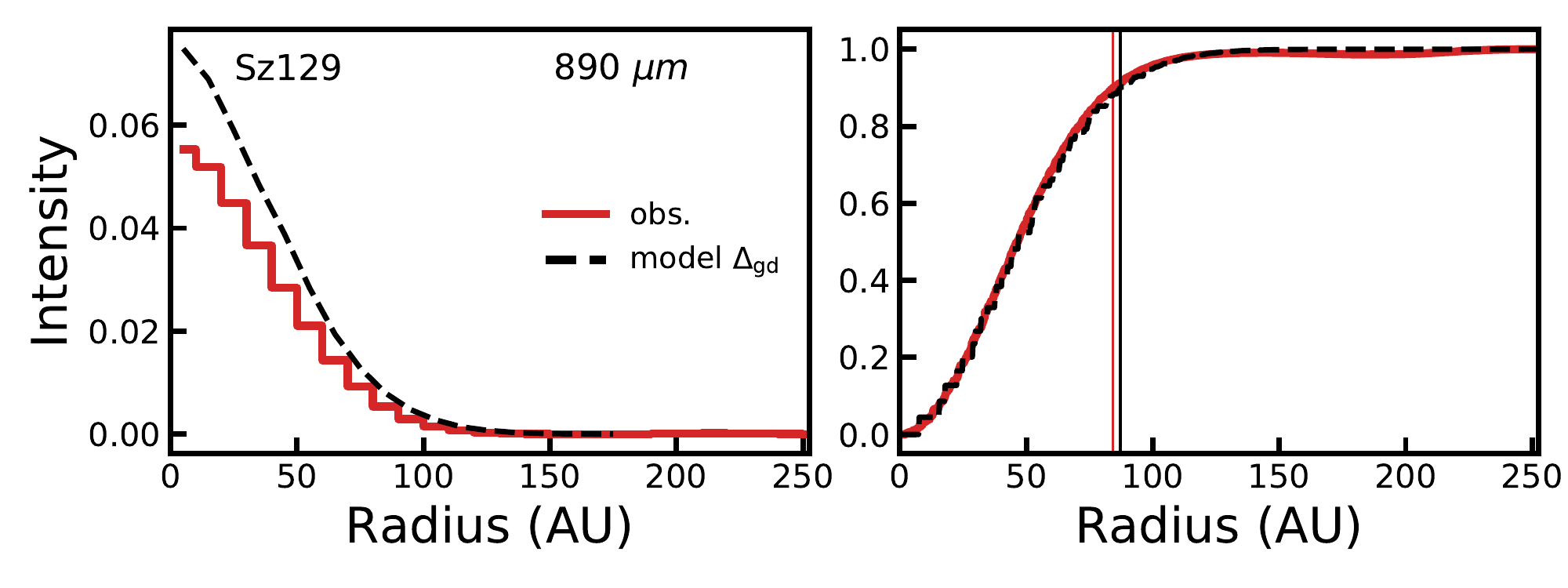}
\end{subfigure}

\begin{subfigure}{0.48\textwidth}
\includegraphics[width=\columnwidth]{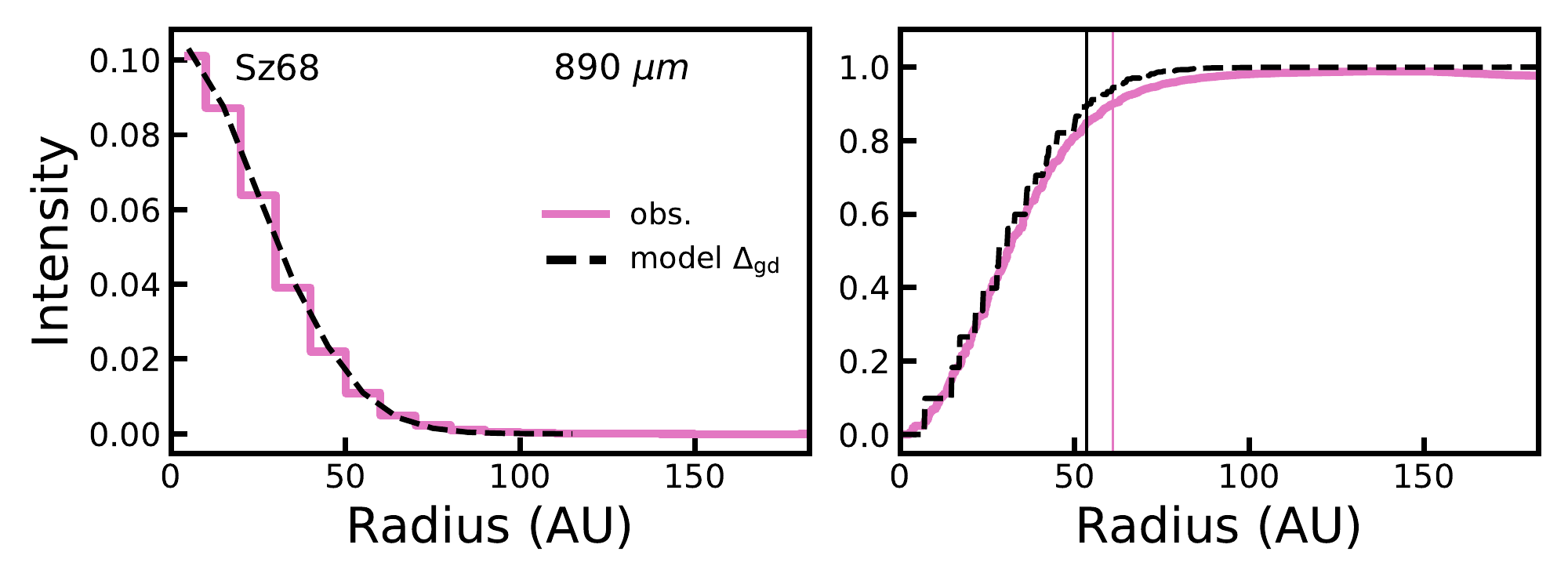}
\end{subfigure}
\begin{subfigure}{0.48\textwidth}
\includegraphics[width=\columnwidth]{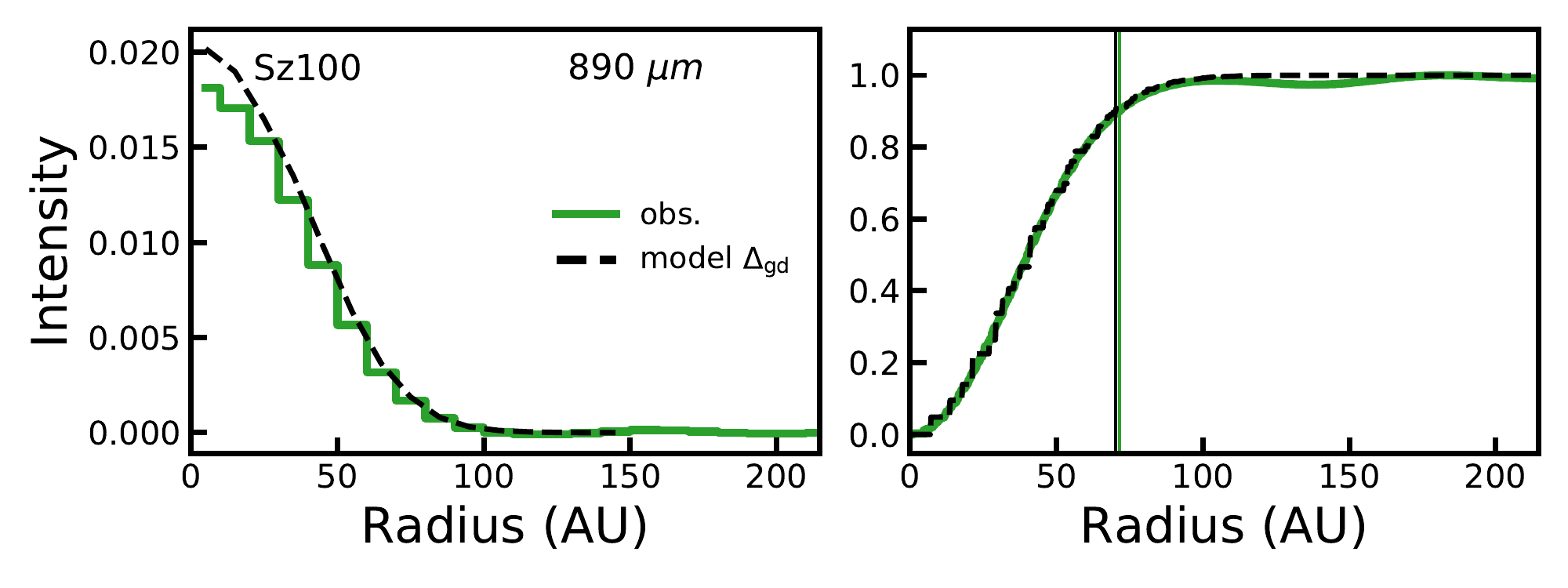}
\end{subfigure}

\begin{subfigure}{0.48\textwidth}
\includegraphics[width=\columnwidth]{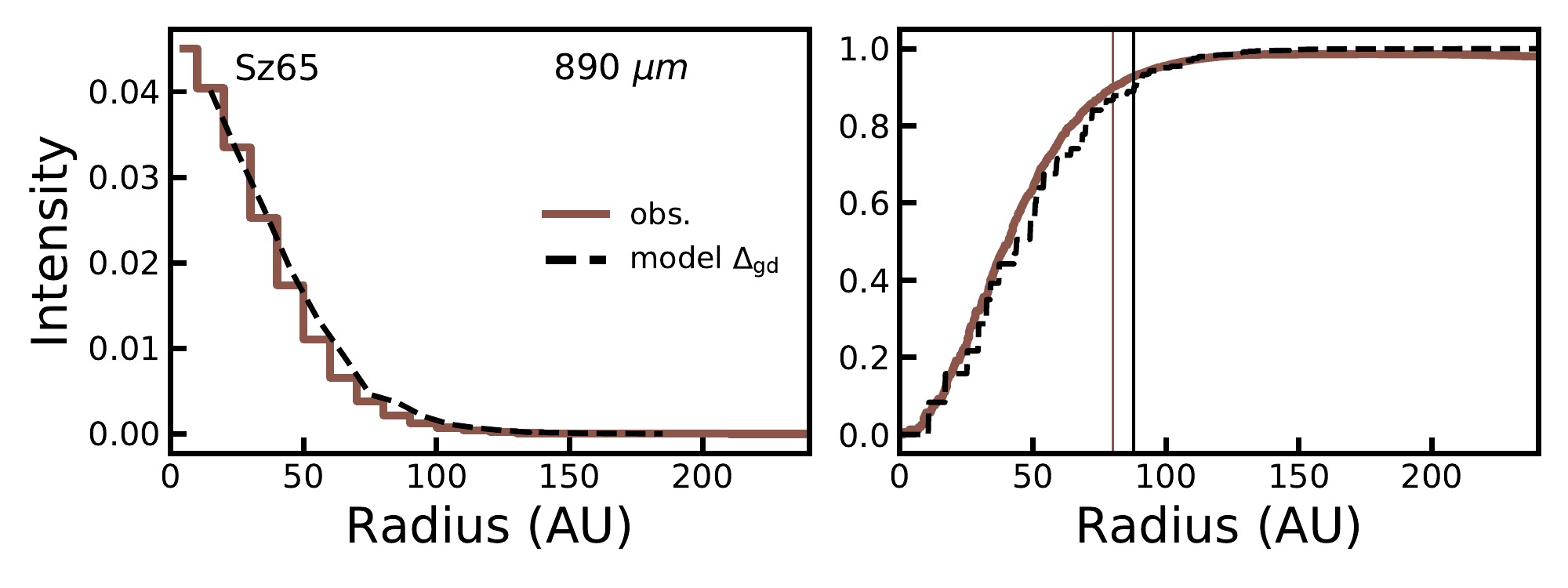}
\end{subfigure}
\begin{subfigure}{0.48\textwidth}
\includegraphics[width=\columnwidth]{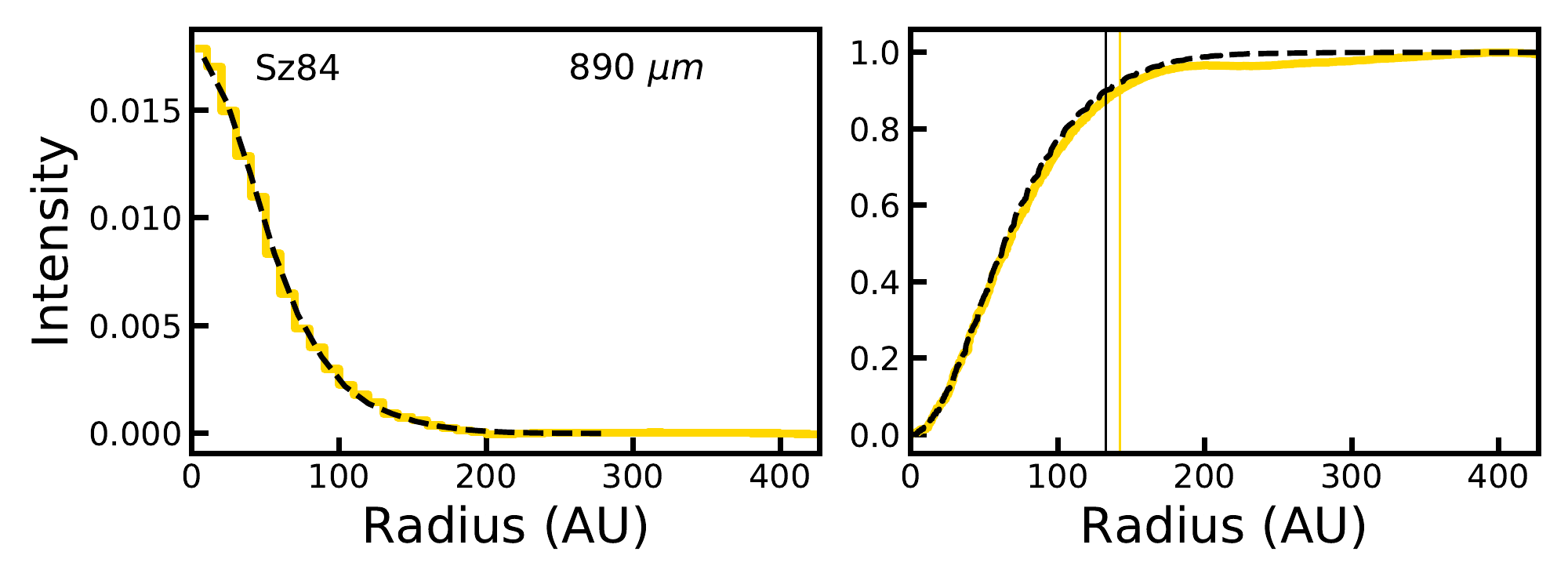}
\end{subfigure}

\caption{\label{fig: 890um continuum comparison} Comparison between model and observed 890~$\mu$m continuum intensity profiles for the ten sources in our sample. For each source, the left panel shows the intensity profile and the right panel shows the matching normalized curve of growth for both the model (black, dashed) and the observation (colored, solid). Vertical lines in the right panels show the radius that encloses 90\% of the total flux (R$_{\rm 890\,\mu m}$). We note that the curve of growth of the model and the observation are normalized separately as the total flux is unimportant when measuring R$_{\rm 890\,\mu m}$. }
\end{figure*}
\FloatBarrier
 
\newpage

\section{Measuring \rgas\ from noisy spectral cube}
\label{app: adding noise to the model}

\begin{figure}[!ht]
    \centering
    \includegraphics[width=\columnwidth]{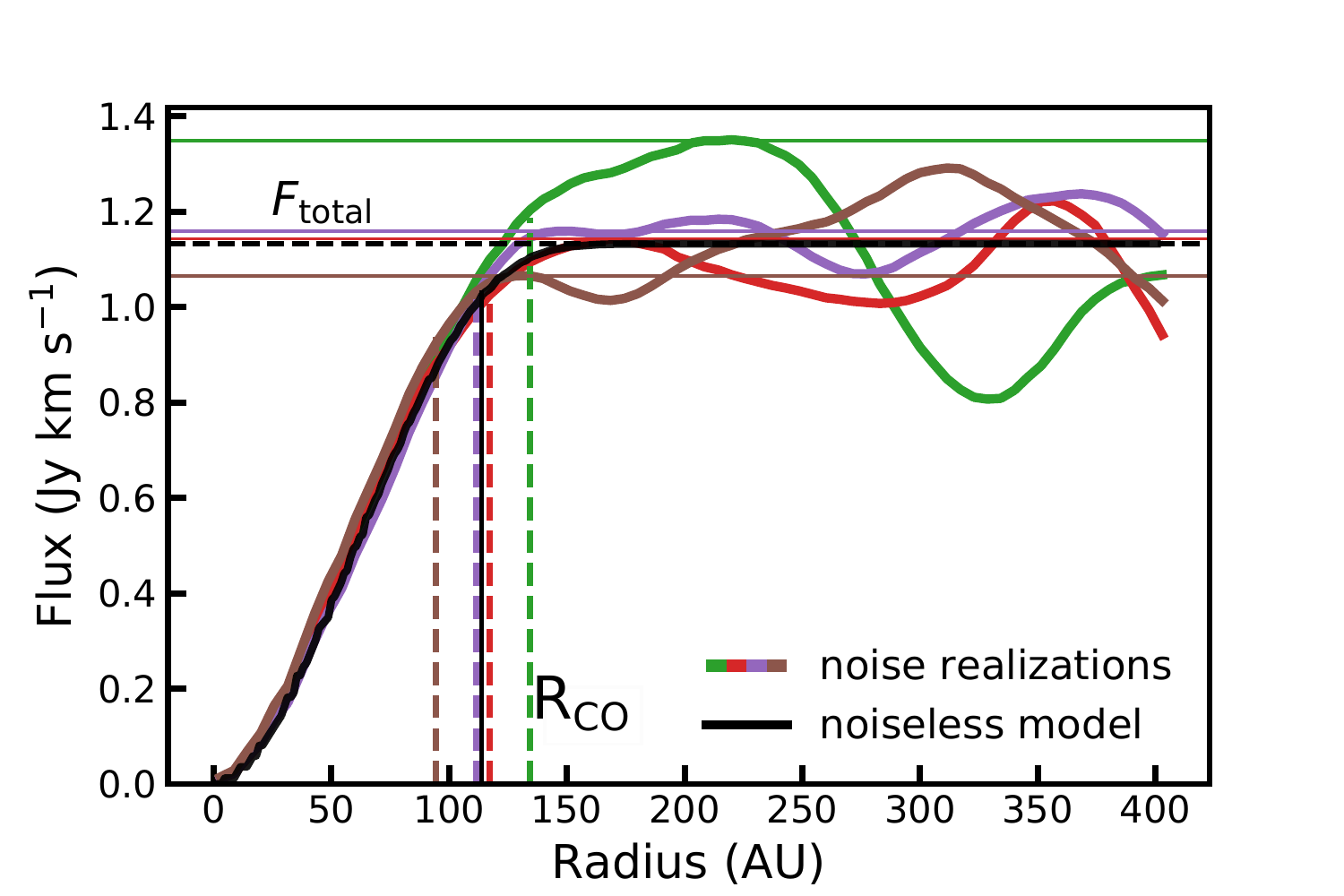}
    \caption{\label{fig: noisy realisation example} Curve of growth of the $^{12}$CO 2\,-\,1 emission for the Sz~129 model for four different and extreme noise realizations chosen to show the range of effects that noise can have on measuring the outer radius. For each noise realization, the horizontal lines show the total flux and the dashed vertical line shows the calculated outer radius. The curve of growth, total flux, and outer radius of the noiseless model are shown in black. }
\end{figure}

Here, we outline how noise was added to models to simulate its effect on measuring \rgas\ from our models. The procedure can be outlined in three steps. First a random patch of noise is added to each of the channels in the synthetic $^{12}$CO spectral cube. The patch of noise is taken from one of the emission free channels in observed spectral cube. Care is taken to avoid the outer edges of the observed field of view in the random selection, as the noise levels there are not representative for the center of the observations.

Next, we construct a Keplerian mask using the same parameters that are used for the observations (cf. Table \ref{tab: mask parameters}). By applying the Keplerian mask to the noisy model spectral cube we increase the S/N in the outer parts of the disk in the same manner as in the observations (cf. Section \ref{sec: measuring model outer radii}). The masked spectral cube is summed over the velocity axis to create the $^{12}$CO moment-zero map. 

Finally, the gas outer radius is measured from the noisy moment-zero map. By definition, \rgas\ is the radius that encloses 90\% of the total flux ($F_{\rm tot}$). In noiseless case, the curve of growth converges to $F_{\rm tot}$ at large radii. In the presence of noise the curve of growth can show variations at large radii, making it difficult to determine $F_{\rm tot}$ (cf. Figure \ref{fig: noisy realisation example}). Empirically, we find that using the flux at the first peak in the curve of growth in most cases most closely matches $F_{\rm tot}$ of the noiseless case. At an average peak S/N$\sim8$ of the $^{12}$CO moment-zero map, noise has a significant effect on measuring \rgas. As an example, Figure \ref{fig: noisy realisation example} shows the curve of growths for four noise realizations for the Sz~129 disk model. As shown in the figure, the curve of growth varies between each noise realization and all four have different $F_{\rm tot}$ and different \rgas. 

\section{Noisy \rgas\ distributions for the disks in our sample}
\label{app: noisy rgas distributions}

\begin{figure*}
\centering
\begin{subfigure}{0.37\textwidth}
\includegraphics[width=\columnwidth]{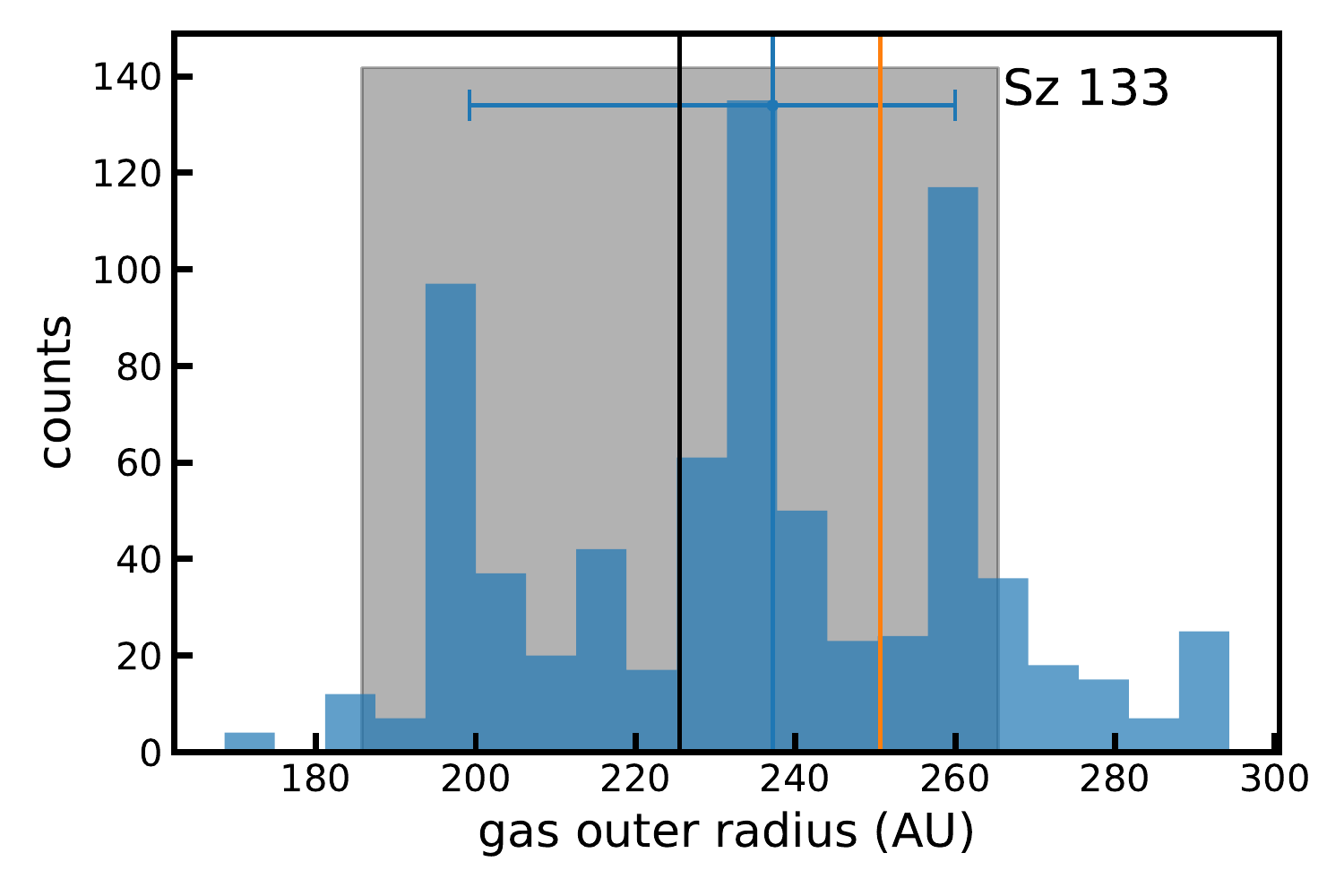}
\end{subfigure}
\begin{subfigure}{0.37\textwidth}
\includegraphics[width=\columnwidth]{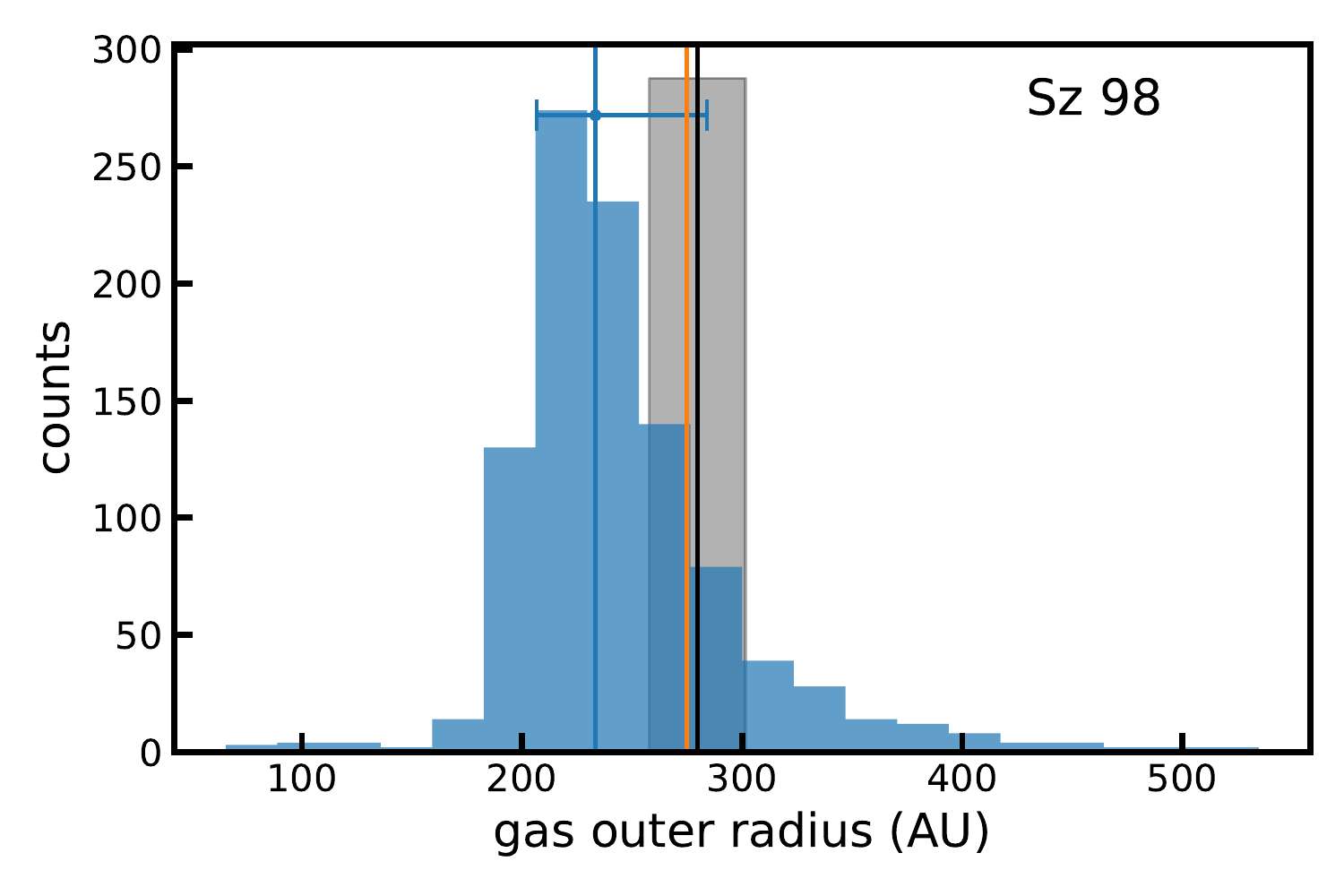}
\end{subfigure}

\begin{subfigure}{0.37\textwidth}
\includegraphics[width=\columnwidth]{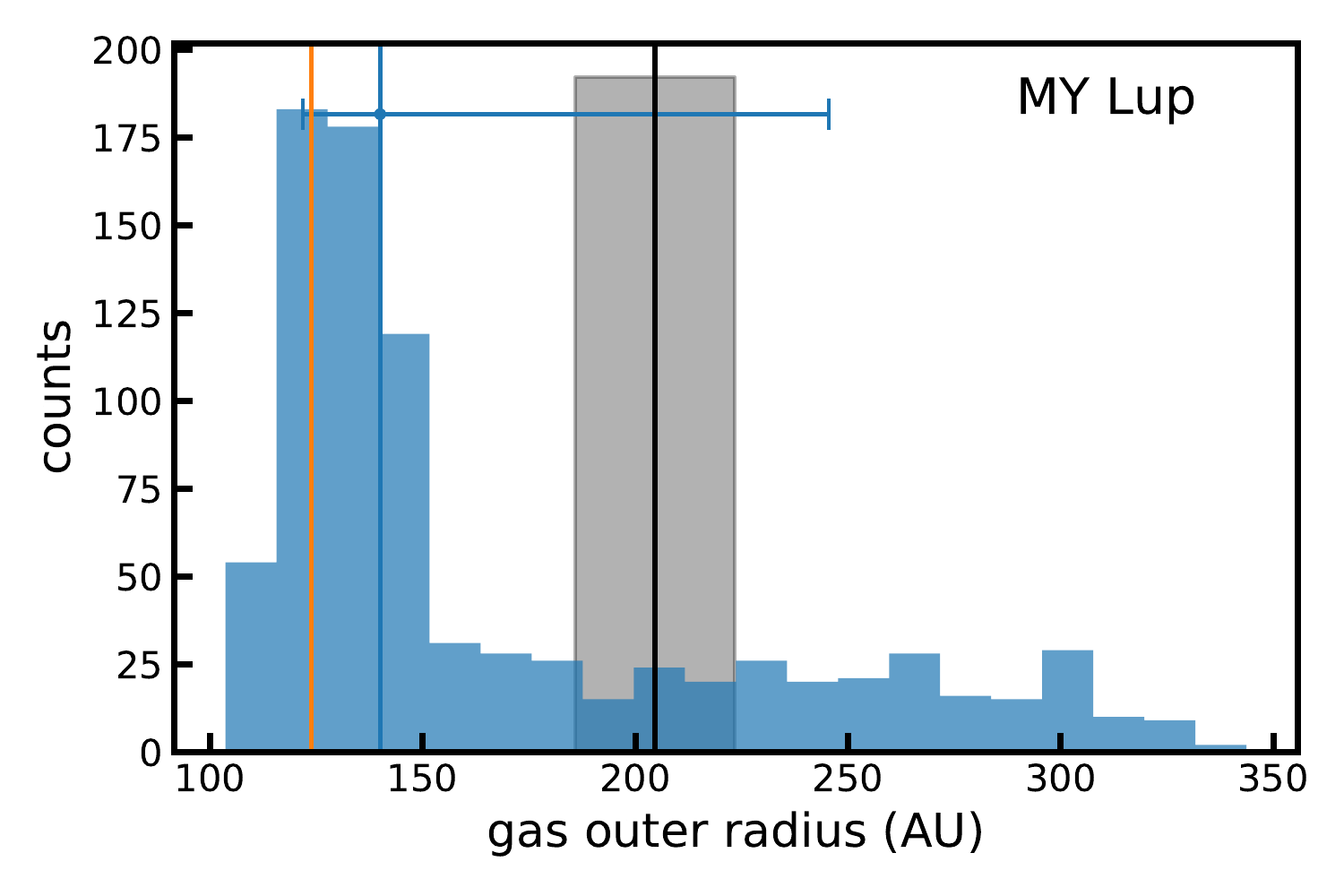}
\end{subfigure}
\begin{subfigure}{0.37\textwidth}
\includegraphics[width=\columnwidth]{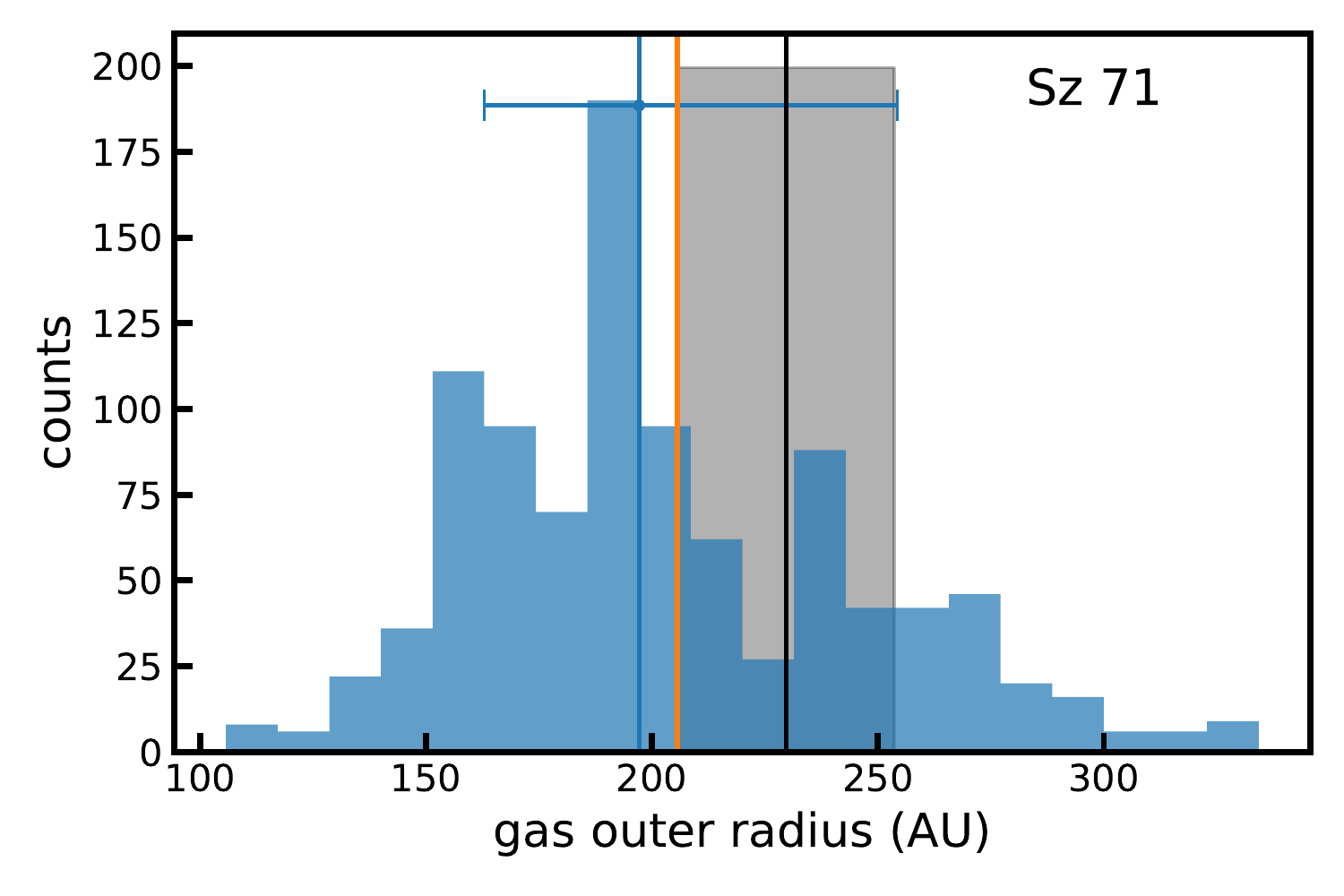}
\end{subfigure}

\begin{subfigure}{0.37\textwidth}
\includegraphics[width=\columnwidth]{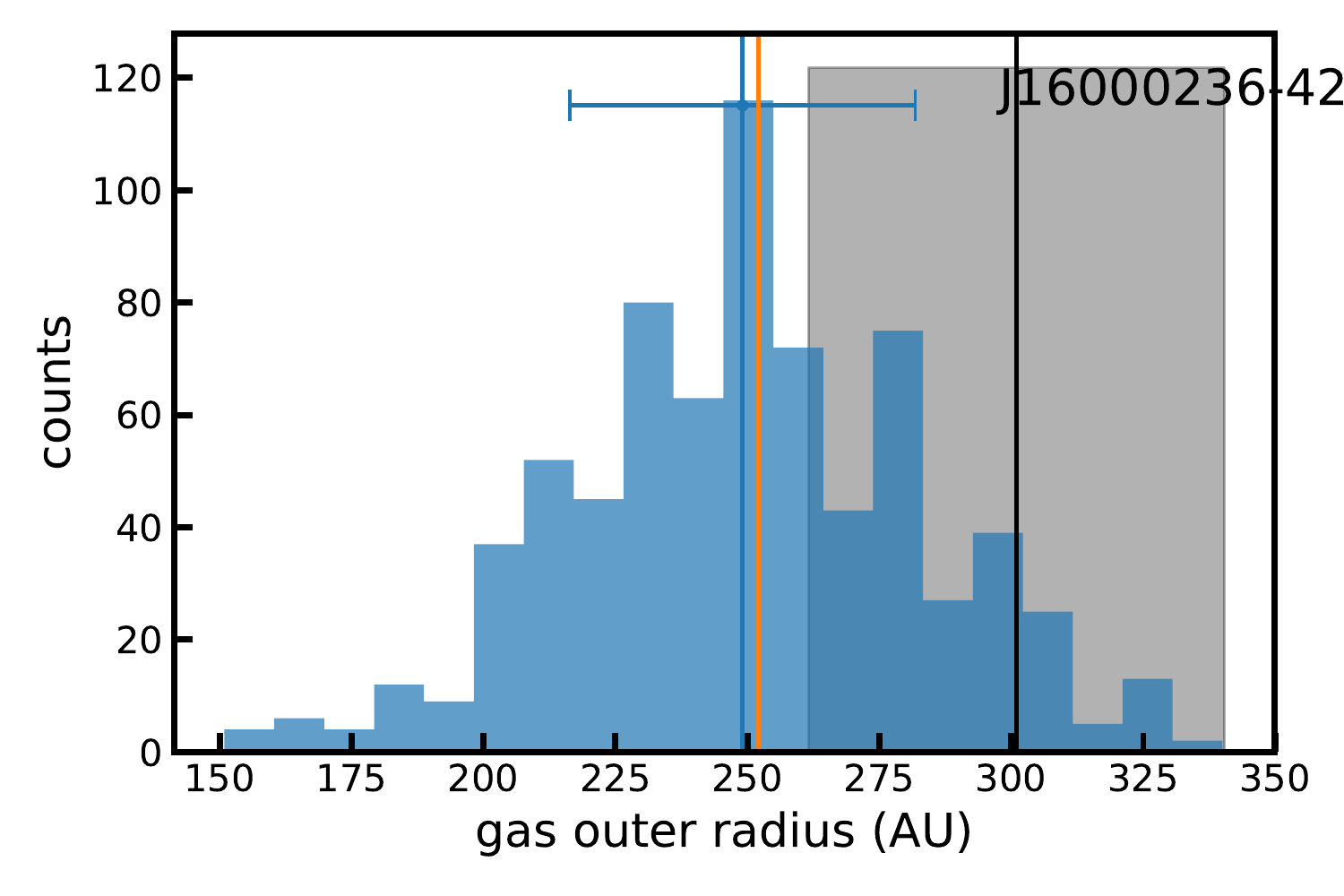}
\end{subfigure}
\begin{subfigure}{0.37\textwidth}
\includegraphics[width=\columnwidth]{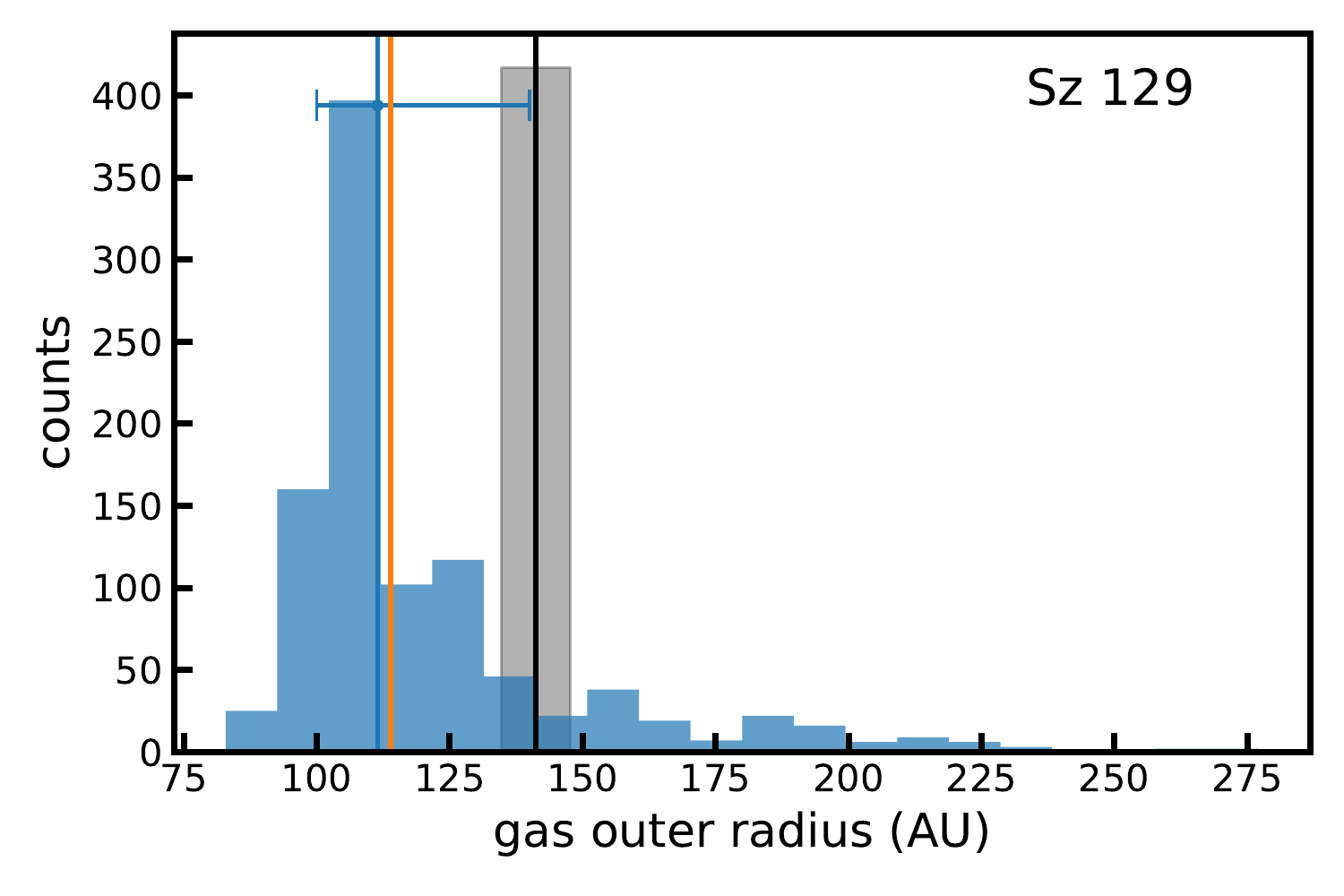}
\end{subfigure}

\begin{subfigure}{0.37\textwidth}
\includegraphics[width=\columnwidth]{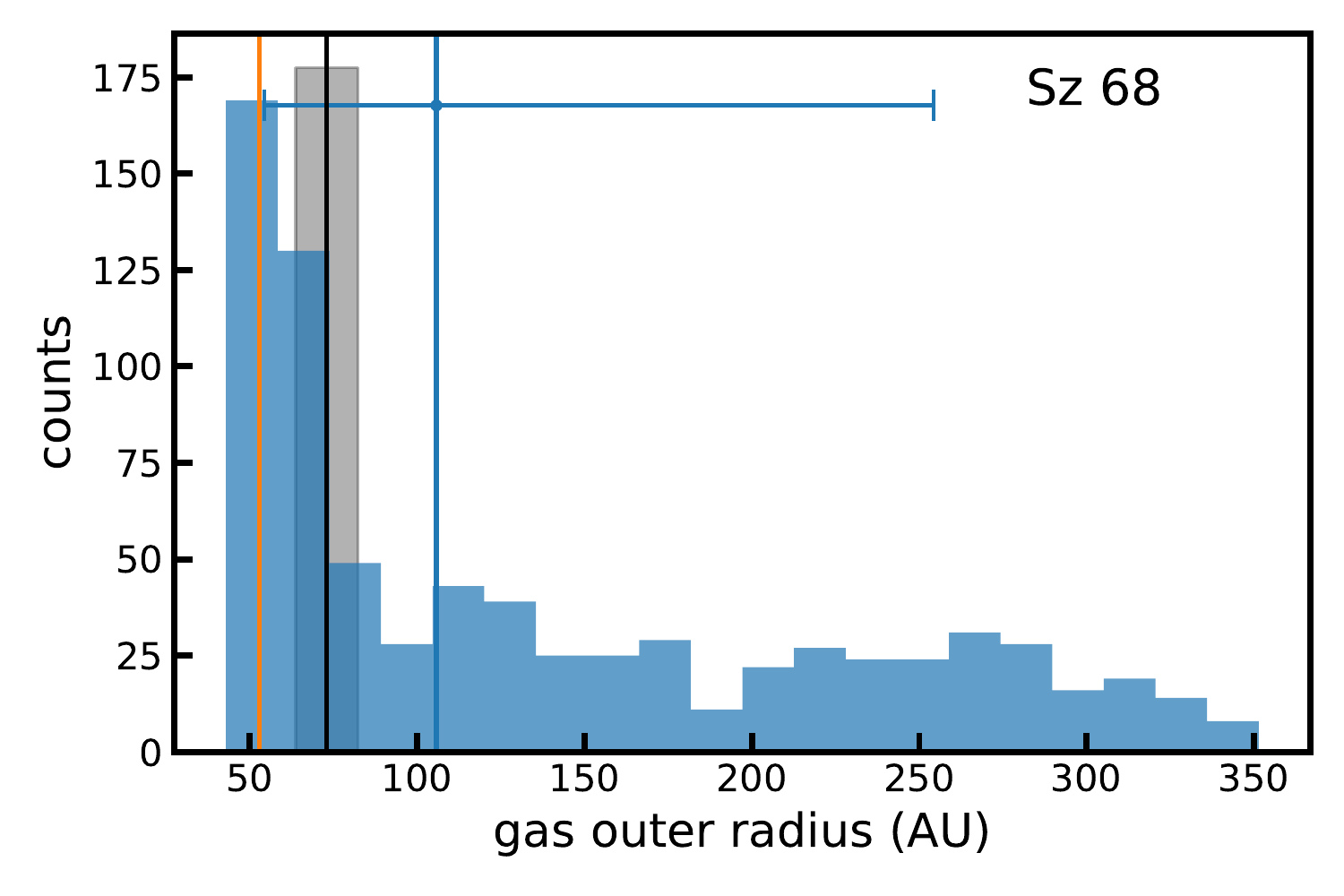}
\end{subfigure}
\begin{subfigure}{0.37\textwidth}
\includegraphics[width=\columnwidth]{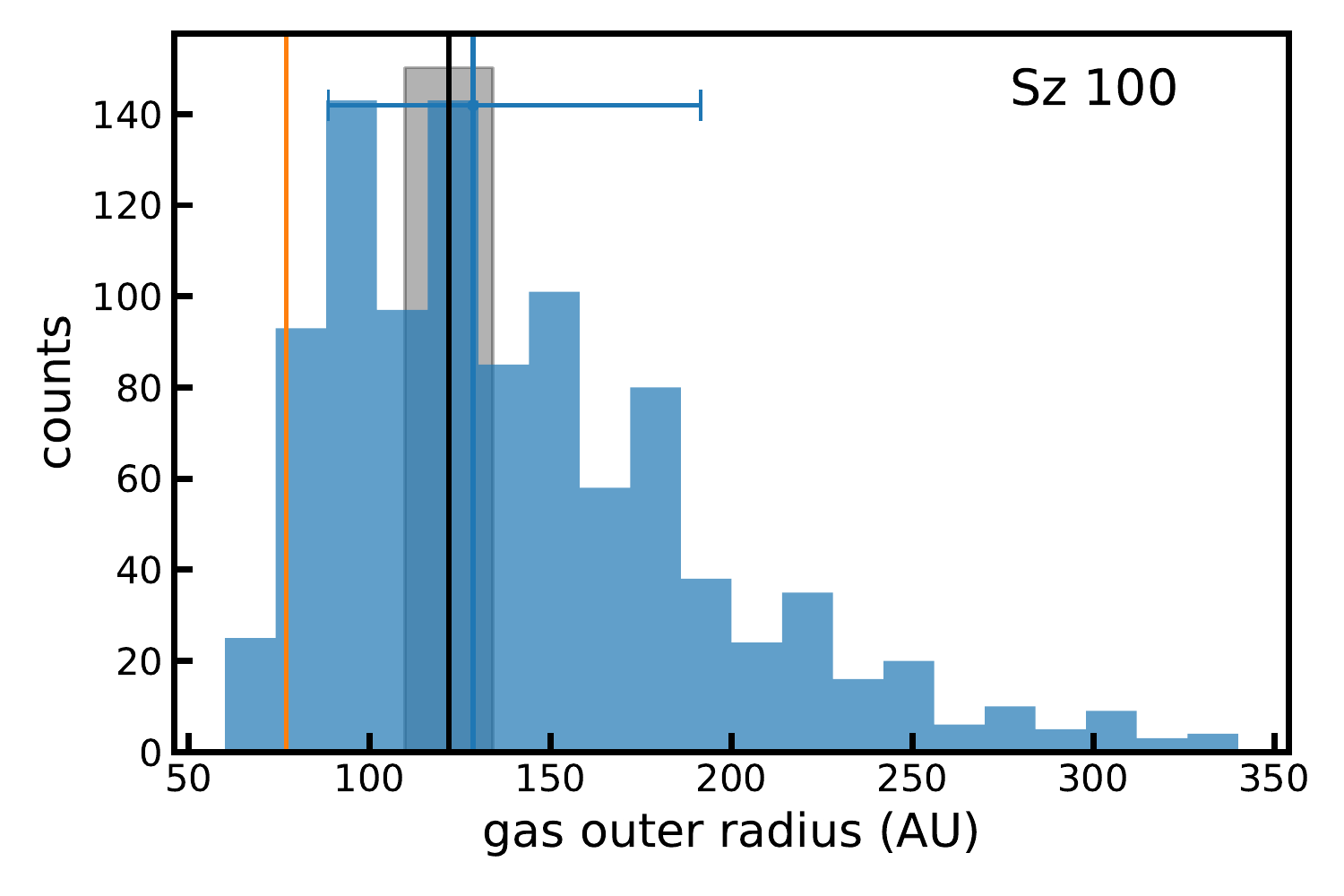}
\end{subfigure}

\begin{subfigure}{0.37\textwidth}
\includegraphics[width=\columnwidth]{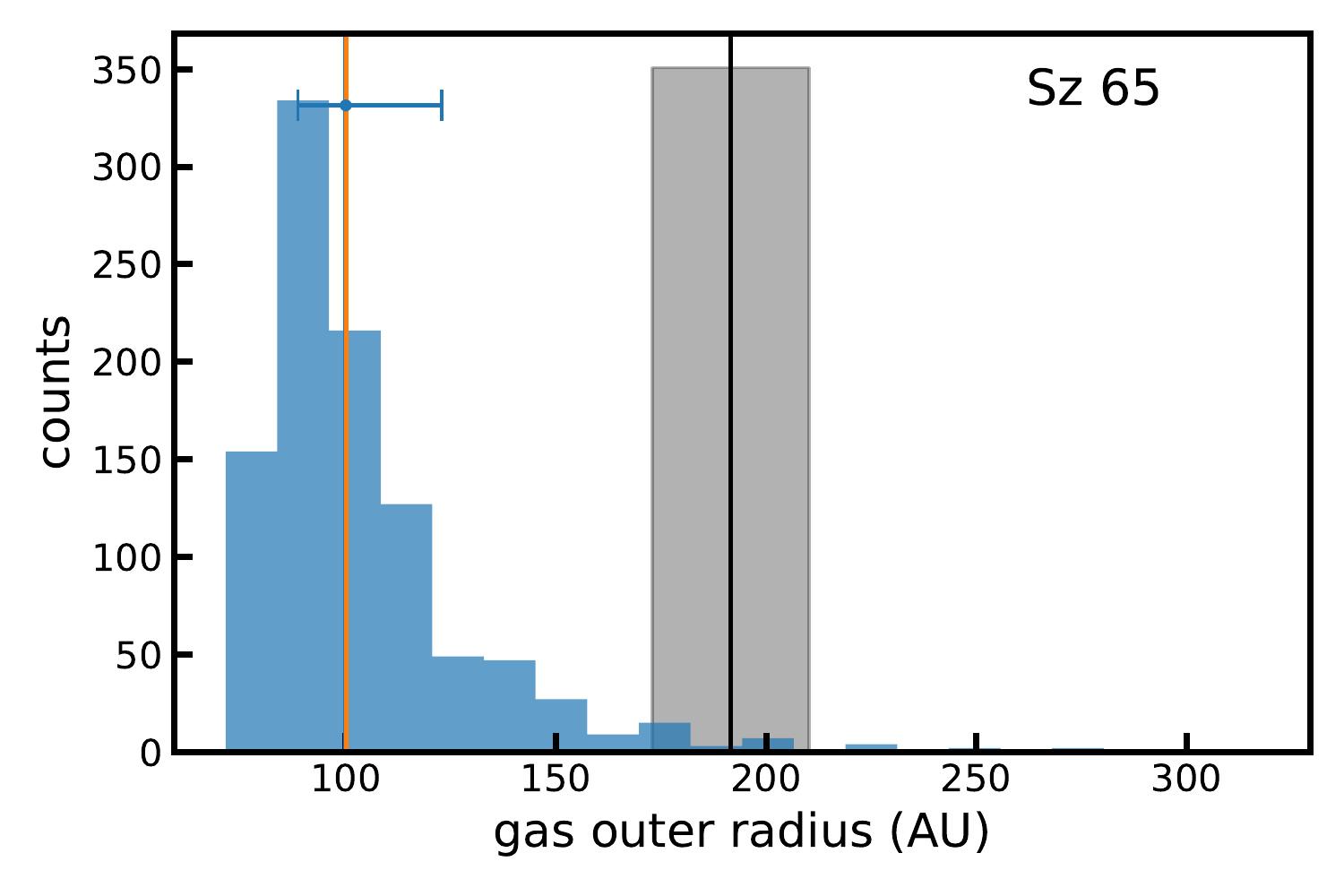}
\end{subfigure}
\begin{subfigure}{0.37\textwidth}
\includegraphics[width=\columnwidth]{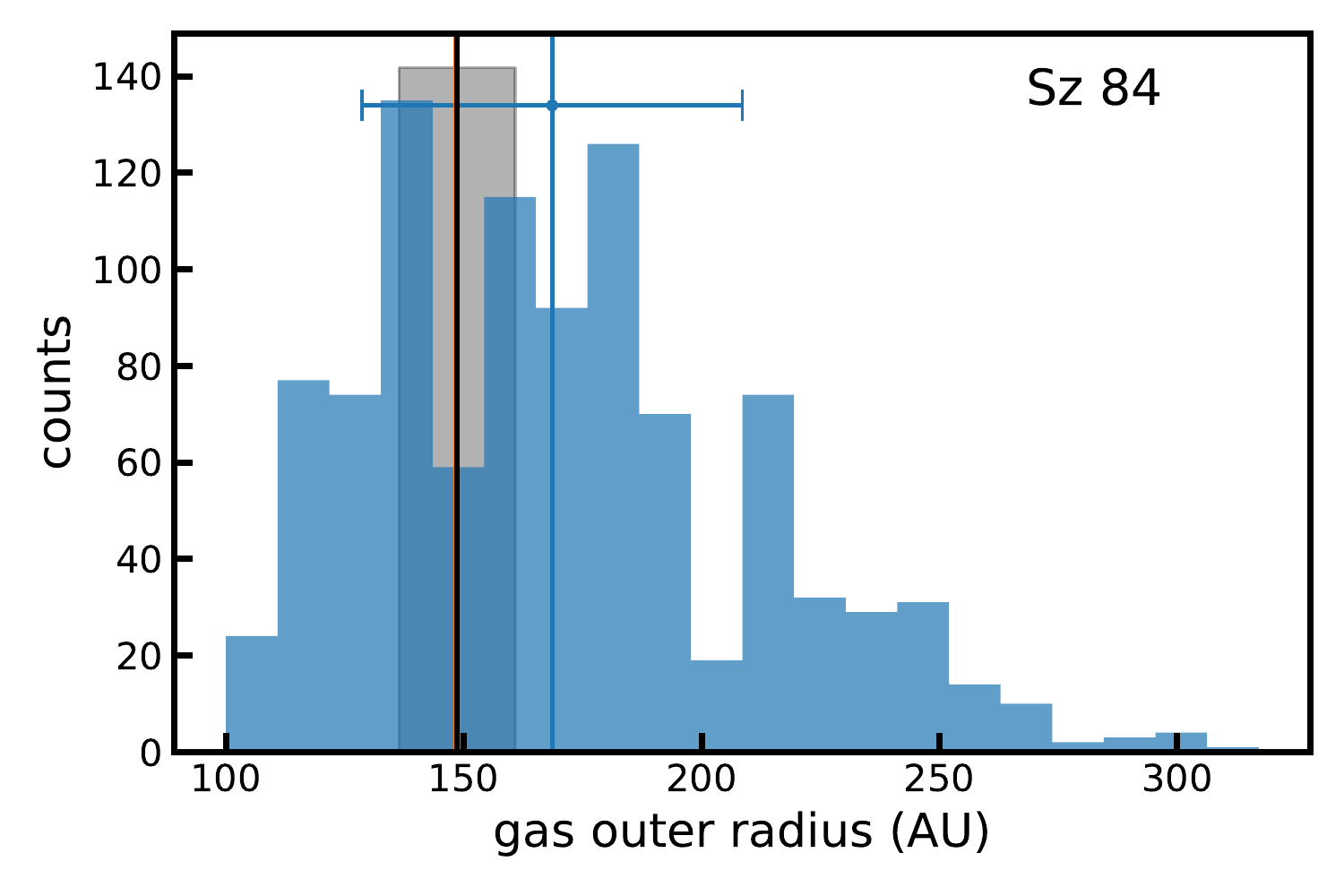}
\end{subfigure}

\caption{\label{fig: noisy rgas realisations - all} Histogram of possible \rgas\ if noise is included in the model. The 16$^{\rm th}$ quantile and the 84$^{\rm th}$ quantile are shown by the blue errorbar, with the median shown as a blue vertical line. Orange vertical line shows \rgas\ of the noiseless model. The observed \rgas\ and uncertainties are shown in black.  }
\end{figure*}

\section{$^{12}$CO $J = 2\,-\,1$ emission maps of 17 sources}
\begin{figure}[!htbp]
     \centering
    \includegraphics[width=\columnwidth]{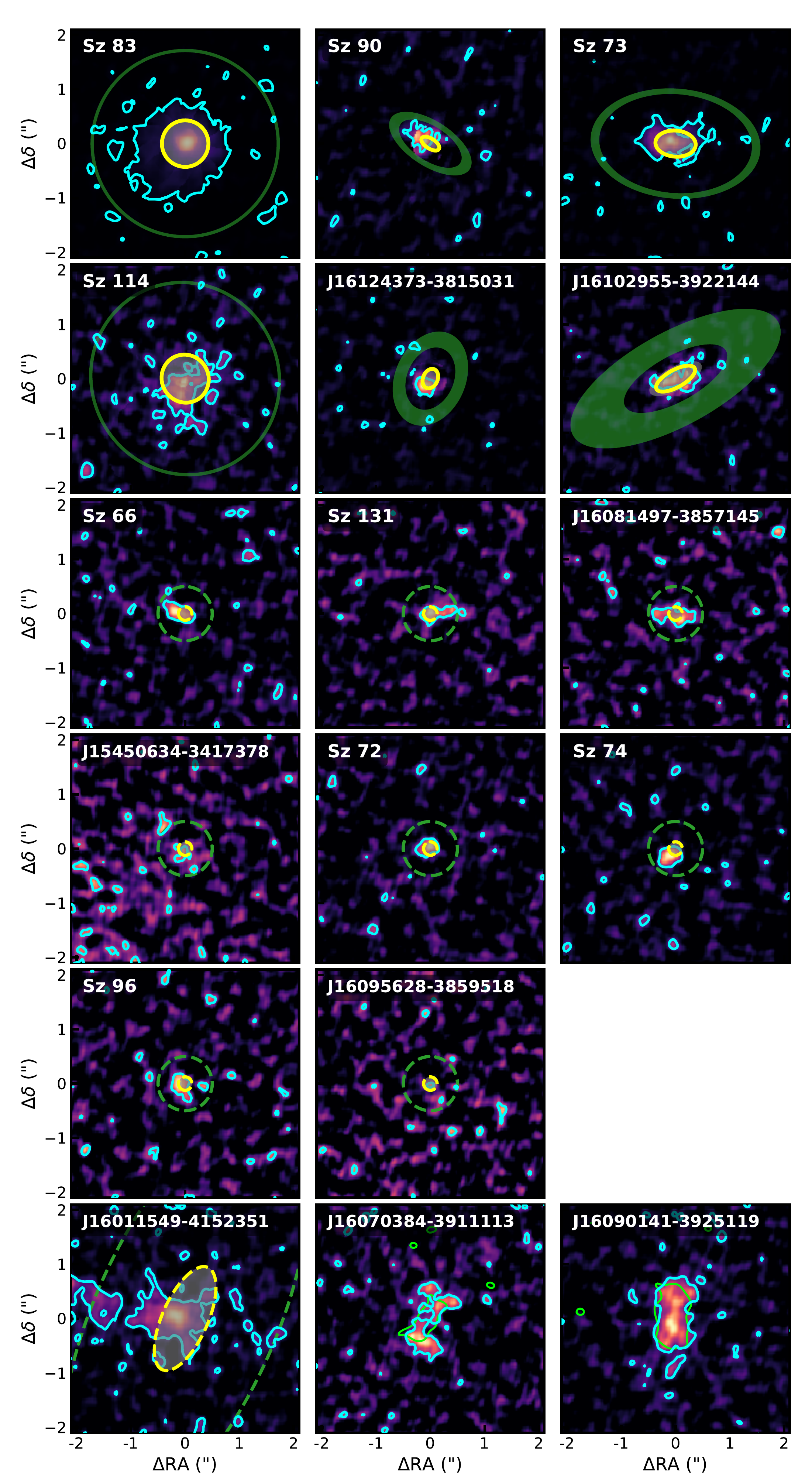}
    \caption{\label{fig: no drift candidates} \CO moment-zero maps of the 17 sources that were detected at low S/N, referred to as the low S/N sample in this work. Cyan contours show significant (S/N $\geq 3$) $^{12}$CO emission. The top six disks have resolved continuum emission and their $\rmmobs$ is shown by the yellow ellipse. 
    For these sources, Keplerian masking was applied to the $^{12}$CO $J = 2\,-\,1$ emission (see Appendix \ref{app: keplerian masking}). 
    The middle eight disks (Sz\,66 to J16095628) have unresolved continuum emission. The dashed yellow circle shows the size of the beam $(0\farcs25)$ as an upper limit to the dust disk size. Similarly, the dashed green circle shows four times the beamsize. 
    For J16011549 the continuum is resolved and R$_{\rm dust}$ was calculated from 90\% of the total flux. The continuum emission of J16070384 and J16090141 is resolved but irregular in shape. Here, we show the $3\sigma$ contours of the continuum in green. }
\end{figure}

In total, 48 disks in Lupus were detected in $^{12}$CO $J = 2\,-\,1$. In this work we analyzed ten of these disks in detail (referred to as the high S/N sample) and showed $^{12}$CO $J = 2\,-\,1$ emission maps of six more disks (part of the low S/N sample; see Section \ref{sec: analyzing remaining disks}). Here we show the $^{12}$CO $J = 2\,-\,1$ emission maps of the remaining disks of the low S/N sample. From our analysis we exclude transition disks with clear resolved cavities. For a detail analysis of these disks, see \cite{vanderMarel2018}. 

The remaining 17 disks with detected $^{12}$CO emission can be divided into two groups. For 6 of the 17 disks the 1.3 millimeter continuum emission was resolved and we were able to measure \rdust\ (see also \citealt{Tazzari2017}). The top two rows of Figure \ref{fig: no drift candidates} show the $^{12}$CO 2\,-\,1 moment-zero maps for these six sources. 
Keplerian masking was applied to the data before making the moment-zero maps (cf. Section \ref{sec: measuring model outer radii} and Appendix \ref{app: keplerian masking}). 

Sz~83, also known as RU~Lup, is a notable inclusion here. This disk is the third-most brightest continuum source in Lupus and has been readily detected in $^{12}$CO, $^{13}$CO, and C$^{18}$O. Furthermore, RU~Lup has also been observed at high spatial resolution as part of the DSHARP program \citep{Andrews2018b}. It seems therefore odd that \rgas\ has not been measured for this source. A closer inspection of the $^{12}$CO channel maps reveals that a component of the emission is nonKeplerian and could be a possible outflow (see Appendix C in \citealt{ansdell2018}). Keplerian masking has removed part of the emission but enough remains in the moment-zero map to prevent us from measuring \rgas. 

There are 11 disks that are detected in $^{12}$CO but for which the 1.3 millimeter continuum emission is not resolved at a resolution of $0\farcs25$.
These are shown in the bottom four rows of Figure \ref{fig: no drift candidates}.
We exclude three disks from our analysis: J16011549\,-\,4152351 is resolved in the continuum, but the $^{12}$CO channel maps show significant cloud emission even after applying Keplerian masking. J16070384\,-\,3911113 and J16090141\,-\,3925119 are resolved but have irregular shaped continuum emission and are possibly unresolved binary sources (cf. \citealt{Tazzari2017}).

\end{appendix}

\end{document}